\documentclass[%
 preprint, 
nofootinbib,
 amsmath,amssymb,
 aps,
pra,
]{revtex4-2}

\usepackage{graphicx}
\usepackage{dcolumn}
\usepackage{bm}
\usepackage[english]{babel}
\usepackage{hyperref}

\usepackage{geometry}

\usepackage{physics}
\usepackage{qcircuit}

\newtheorem{definition}{Definition}[section]

\newtheorem{lemma}{Lemma}[section]
\newtheorem{proposition}{Proposition}[section]

\newtheorem{remark}{Remark}[section]
\newtheorem{theorem}{Theorem}[section]
\newcommand{\cqd}{\hfill $\rule{2.5mm}{2.5mm}$ \vspace{0.4cm}}

\begin{document}

\preprint{APS/123-QED}

\title{\textbf{Quantum Simulation of Quantum Channels and Applications.} 
}%

\author{Lúcio S. Fassarella}
\affiliation{%
Departamento de Matemática Aplicada, Universidade Federal do Espírito Santo, São Mateus, ES, Brazil
}%

%

\date{\today}

\begin{abstract}
 
\end{abstract}

\maketitle


\section{\label{secao_intro}Introduction}

\emph{Quantum channels}%
\footnote{
	There are terms in the literature denoting notions equivalent or closely related to the concept of a \emph{quantum channel}, such as: \emph{dynamical map} \cite{Breuer2002}, \emph{quantum map} \cite{Lidar2020}, and \emph{quantum operation} \cite{Nielsen2010, Olivares2025}. In particular, \emph{quantum operations} are transformations that generalize quantum channels and serve to describe processes in quantum systems that do not necessarily preserve the trace. It is worth noting also that, despite their generality, quantum channels do not describe all processes that can occur in quantum states \cite{Shaji2005}.}%
	 are mathematical objects that describe very general transformations on the states of quantum system, encompassing \emph{positive operator-valued measures} (POVMs), \emph{decoherence}, \emph{noise}, and the evolution of \emph{Markovian quantum systems} \cite{Alicki1987, Bengtsson2017, Breuer2002, Nielsen2010, Olivares2025, Pinto2023, Watrous2018, Wilde2016}. Given the variety of applications and potential scenarios where the analytical description of quantum channels requires an excessive or even impossible amount of manual calculation, it is natural to seek strategies for their computational implementation. When considering the simulation of quantum channels on \emph{digital quantum computers}\footnote{Digital quantum computers operate according to the so-called \emph{circuit model}, in which a discrete, finite number of \emph{qubits} is prepared and manipulated via \emph{quantum logic gates} (unitary operators) and its state ends up being measured to provide useful information. A classic, concise description of quantum computing within the circuit model is provided in \cite[Section 4.6, p.202]{Nielsen2010}.
}%
, we encounter an initial difficulty: quantum channels are not necessarily unitary, whereas digital quantum computers manipulate quantum states exclusively through unitary operations (referred to in this context as \emph{quantum gates}) \cite{Nielsen2010, Olivares2025, Watrous2018, Wilde2016}. The objective of this article is to demonstrate how to circumvent this difficulty by presenting a strategy for simulating quantum channels based on the \emph{Choi-Kraus operator-sum representation} -- we term \emph{UD strategy} (short for \emph{unitary dilation}) --, which allows us to address various applications in a unified manner, including the preparation of mixed states, operator-valued measurements, and the evolution of Markovian quantum systems. Although this subject or aspect of it are addressed in many other works, \textit{e.g.}, \cite{Accardi1999, Cleve2019, Gohm2024, Guan2024, Pinto2023, Ying2021, Wang2026, Ziman2005}, we believe we offer a contribution through the unified treatment of the aforementioned topics which is suitable for digital quantum computation 

Focusing on quantum simulations of the evolution of Markovian quantum systems, our approach follows the one presented in the first part of the Introduction in \cite{Cleve2019} and also in \cite{Ding2024}; however, we introduce a new element -- the so-called \emph{residual operator} -- which allows us to frame the problem within the scope of the UD strategy for the quantum simulation of quantum channels, without resorting to \emph{Trotterization} \cite{Nielsen2010}, \emph{stochastic differential equations} \cite{Ding2024}, or other methods.This \emph{residual operator} allows us to construct, directly from the \emph{Liouvillian}, a quantum channel that approximates the system's evolution operator to first order, thereby enabling a quantum simulation analogous to the Euler method for ordinary differential equations 

In Section \ref{secao_canaisq} we review basic concepts and results concerning quantum channels and develop the UD strategy for simulating them on digital quantum computers, in addition to establishing the notation. We present three applications: preparation of mixed states (Section \ref{secao_prepest}), simulation of POVM measurements (Section \ref{secao_povm}) and quantum computational simulation of the evolution of Markovian quantum systems (Section \ref{secao_sistemas-markovianos}). In Section \ref{secao_conclusao}, we review the results and offer some concluding remarks 


\section{\label{secao_canaisq}Quantum Channels and UD strategy for Quantum Simulation}

We recall that in quantum mechanics, every physical system is associated with a Hilbert space, and its observables and states are described, respectively, by Hermitian operators and density operators on that Hilbert space \cite{Nielsen2010, Olivares2025, Wilde2016} 

Given finite dimensional Hilbert spaces $\mathcal{H}$, we denote:
\begin{itemize}
	\item  $I_\mathcal{H}$ the identity operator on $\mathcal{H}$;
	\item $\mathcal{L}\qty(\mathcal{H})$ the space of linear operators from $\mathcal{H}$ to $\mathcal{H}$; 
	\item $\mathcal{D}\qty(\mathcal{H}) \subset \mathcal{L}\qty(\mathcal{H})$ the convex set of density operators on $\mathcal{H}$ (positive Hermitian operators with unit trace):
	\[
	\mathcal{D}\qty(\mathcal{H}) = \qty{\rho \in \mathcal{L}\qty(\mathcal{H})/\ \rho^\dagger = \rho,\ \rho \ge 0,\ \tr(\rho) = 1}. 
	\]
	\item Notation for the tensor product of state vectors:
	\[
	\ket{\psi \otimes \phi} \equiv \ket{\psi}\otimes \ket{\phi},\ \forall \ket{\psi}\in \mathcal{H},\ \forall \ket{\psi}\in \mathcal{H}'.
	\]
\end{itemize}

\begin{definition}[Quantum channel]\label{def_canal-quantico}
	A quantum channel from $\mathcal{H}$ to $\mathcal{H}'$ is a linear map $\mathcal{E}: \mathcal{L}(\mathcal{H}) \rightarrow \mathcal{L}(\mathcal{H}')$ that satisfies the following pair of conditions:%
	\footnote{%
		There is no physical motivation to require a quantum channel to be defined on all linear operators of the Hilbert space; it suffices for it to be defined on density operators and be \emph{convex}, \textit{i.e.},
		\[
		\mathcal{E}\qty[s\rho + (1-s)\sigma] = s\mathcal{E}\qty[\rho] + (1-s)\mathcal{E}\qty[\sigma],\ \forall \rho,\sigma \in \mathcal{D},\ \forall s \in \qty[0,1]; 
		\]
		however, the linearity condition entails no loss of generality because it can be proven that any convex map on the set of density operators that is also completely positive and trace-preserving can be extended to a linear operator on the entire Hilbert space that also possesses these properties \cite[Appendix B]{Wilde2016}. 
	}%
	\begin{itemize}
		\item[i)] $\mathcal{E}$ is \emph{trace-preserving}:
		\[
		\tr\qty(\mathcal{E}\qty[\Gamma]) = \tr(\Gamma),\ \forall \Gamma \in \mathcal{L}(\mathcal{H}). 
		\]
		\item[ii)] $\mathcal{E}$ is \emph{completely positive}, \emph{i.e.},: given any finite-dimensional Hilbert space $\mathcal{K}$, the operator
		\[
		\qty(I_\mathcal{K}\otimes \mathcal{E}) : \mathcal{L}\qty(\mathcal{K}\otimes \mathcal{H}) \rightarrow \mathcal{L}\qty(\mathcal{K}\otimes \mathcal{H}')
		\]
		maps positive-semidefinite operators to positive-semidefinite operators, that is,
		\[
		\qty(I_\mathcal{K}\otimes\mathcal{E}) \qty[\Gamma] \ge 0,\ \forall \Gamma \in \mathcal{L}\qty(\mathcal{K}\otimes \mathcal{H}),\ \Gamma \ge 0.
		\]
		\begin{itemize}
			\item[$\bullet$] The quantum channel is \emph{unitary} when there exists a unitary operator $U: \mathcal{H}\rightarrow \mathcal{H}'$ such that
			\[
			\mathcal{E}\qty[\Gamma] = U\Gamma U^\dagger,\ \forall \Gamma \in \mathcal{L}(\mathcal{H}).
			\]
			\item[$\bullet$] The quantum channel is \emph{unital} when it associates identity operator to identity operator:
			\[
			\mathcal{E}\qty[I_\mathcal{H}] = I_{\mathcal{H}'}.
			\]
		\end{itemize}
	\end{itemize}
\end{definition}

\begin{remark}
	In this definition, it is assumed that the \emph{complete positivity} of a channel $\mathcal{E}$ implies that it is also \emph{positive}:
	\[
	\mathcal{E} \qty[\Gamma] \ge 0,\ \forall \Gamma \in \mathcal{L}\qty(\mathcal{H}),\ \Gamma \ge 0.
	\]
\end{remark}

For physical interpretations of the conditions characterizing quantum channels in Definition \ref{def_canal-quantico}, see \cite{Nielsen2010} or \cite{Wilde2016}. Fortunately, we will not need to deal directly with the complete positivity of quantum channels, other than recalling that this property is equivalent to the existence of a \emph{Choi-Kraus representation} (Theorem \ref{teo_kraus} below).\footnote{%
	The characterization of completely positive operators can be reduced to the characterization of positive operators via the \emph{Choi-Jamiołkowski representation} \cite[Proposition 2.1, p.33]{Wolf2012}: \emph{A linear map $T: \mathcal{L}\qty{\mathcal{H}} \rightarrow \mathcal{L}\qty{\mathcal{H}'}$ is completely positive if and only if $\qty(T\otimes I_\mathcal{H})\qty(\ketbra{\Omega}) \ge 0$, where $\Omega$ is a maximally entangled state of dimension $d=\dim \mathcal{H}$.} The operator $T \otimes I_\mathcal{H}$ is called the \emph{Jamiołkowski state}. Notably, this characterization is related to the \emph{channel-state duality}, a correspondence relating dynamics to kinematics that ``[...] is not merely mathematical but also has a fundamental physical meaning, profound consequences, and a plethora of applications.'' \cite[our translation]{Jiang2013}.
}

\begin{theorem}[Choi-Kraus representation]\label{teo_kraus} A linear map $\mathcal{E}: \mathcal{L}\qty(\mathcal{H}) \rightarrow \mathcal{L}\qty(\mathcal{H}')$ is a quantum channel if, and only if, there exists a finite family of so-called \emph{Kraus operators}
\[
\qty{E_k: \mathcal{H} \rightarrow \mathcal{H}';\ k = 1, \dots, m}
\] 
such that
	\begin{equation}\label{choi-kraus_representation}
	\mathcal{E}\qty[\Gamma] = \sum_{k=1}^{m} E_k\Gamma E_k^\dagger,\ \forall \Gamma \in \mathcal{L}\qty(\mathcal{H}')
	\end{equation}
	and\footnote{
		This second identity is equivalent to trace preservation condition of Definition \ref{def_canal-quantico}.
	}
	\begin{equation}\label{choi-kraus_isometry}
	\sum_{k=1}^{m}E_k^\dagger E_k = I.
	\end{equation}
	Furthermore, the order of the Choi-Kraus representation can be limited by the product of the dimensions of $\mathcal{H}$ and $\mathcal{H}'$:
	\[
	m \le \dim\qty(\mathcal{H})\dim\qty(\mathcal{H}'). 
	\]
\end{theorem}
For proofs of Theorem \ref{teo_kraus}, see \cite{Choi1975, Kraus1983, Nielsen2010, Watrous2018, Wilde2016} 

In the simulation of quantum channels on digital quantum computers, we will employ two objects constructed from a Choi-Kraus representation: the \emph{Stinespring dilation} and (what we term) the \emph{unitary dilation}. In essence, the unitary dilation extends the quantum channel to the tensor product of the original Hilbert space and an ancilla Hilbert space, often referred to as the \emph{environment}, in such a way that the quantum channel can be recovered by taking the partial trace over the ancilla system.\footnote{%
	For a precise definition of the partial trace and some related identities, see \cite[Section 4.3.3]{Heinosaari2011}.
}

\begin{definition}[Stinespring dilation]\label{def_ditatacao-stinespring}
	Consider a quantum channel $\mathcal{E}: \mathcal{L}\qty(\mathcal{H}) \rightarrow \mathcal{L}\qty(\mathcal{H}')$. A \emph{Stinespring dilation} of $\mathcal{E}$ with ancilla Hilbert space $\mathcal{G}$ is a linear map $V: \mathcal{H} \rightarrow \mathcal{H}' \otimes \mathcal{G}$ such that
	\begin{equation}\label{eq_stinespringdilation}
	\mathcal{E}\qty[\Gamma] = \tr_\mathcal{G}\qty(V\Gamma V^\dagger),\ \forall \Gamma \in \mathcal{L}\qty(\mathcal{H}),
	\end{equation}
	and
	\begin{equation}\label{eq_stinespringunitarity}
	V^\dagger V = I.
	\end{equation}
\end{definition}

\begin{remark}\label{obs_dilatacao-stinespring}
	Highlights regarding Definition \ref{def_ditatacao-stinespring}:\footnote{
		For further details on the Stinespring dilation, see \cite{Bengtsson2017, Paulsen2003, Stinespring1955, Watrous2018, Wilde2016}. 
	}
	\begin{itemize}
		\item[i)] The identity $\mathcal{E}\qty[\Gamma] = \tr_\mathcal{G}(V\Gamma V^\dagger)$ for all $\Gamma \in \mathcal{L}\qty(\mathcal{H})$ is equivalent to this condition restricted to rank-one projections on $\mathcal{H}$, since $\mathcal{E}$ is linear on $\mathcal{L}\qty(\mathcal{H})$ and any linear operator on $\mathcal{H}$ can be written as a linear combination of rank-one projections; that is,
		\[
		\begin{split}
			\mathcal{E}\qty[\Gamma] &= \tr_\mathcal{G}\qty(V\Gamma V^\dagger),\ \forall \Gamma \in \mathcal{L}\qty(\mathcal{H}) \iff\\
			\mathcal{E}\qty[\ketbra{\psi}] &= \tr_\mathcal{G}\qty(V\ketbra{\psi} V^\dagger),\ \forall \ket{\psi} \in \mathcal{H}.
		\end{split}
		\]
		\item[ii)] The identity $V^\dagger V = I$ (isometry condition) is equivalent to $\mathcal{E}$ being trace-preserving.
	\end{itemize}
\end{remark}

\begin{theorem}[Stinespring dilation from Choi-Kraus representation]\label{teo_dilatacao-stinespring}
	Let $\mathcal{E}: \mathcal{L}\qty(\mathcal{H}) \rightarrow \mathcal{L}\qty(\mathcal{H}')$ be a quantum channel and $\qty{E_k;\ k = 1, \dots, m}$ be a Choi-Kraus representation of $\mathcal{E}$ of order $m\in \mathbb{N}$. Define $\mathcal{G}:= \mathbb{C}^m$ and fix an orthonormal basis for $\mathcal{G}$, say $\qty{\ket{u_k};\ k = 1, \dots, m}$. Then, the following map is a Stinespring dilation of $\mathcal{E}$:
	\[
	V: \mathcal{H} \rightarrow \mathcal{H}' \otimes \mathcal{G},\ V \ket{\psi} := \sum_{k=1}^{m} E_k\ket{\psi}\otimes \ket{u_k},\ \forall \ket{\psi}\in \mathcal{H}. 
	\]
	In addition to $V$ being an isometry, $VV^\dagger$ is an orthogonal projection on $\mathcal{H}'\otimes \mathcal{G}$, \textit{i.e.}, $VV^\dagger$ is Hermitian and idempotent:
	\[
	(VV^\dagger)^2 = VV^\dagger. 
	\]
	\begin{proof}
		We prove each item of Theorem \ref{teo_dilatacao-stinespring} by direct inspection. First, we observe that the Hermitian conjugate of $V$ is given by
		\[
		V^\dagger(\cdot) = \sum_{k=1}^m E_k^\dagger\qty(\cdot) \otimes \bra{u_k};
		\]
		so, for a given $\Gamma \in \mathcal{L}\qty(\mathcal{H}_A)$, it holds
		\[
		V\Gamma V^\dagger = \sum_{j=1}^m\sum_{i=1}^m  E_j\Gamma E_i^\dagger \otimes \ketbra{u_j}{u_i}.
		\]
		From this, we get
		\[
		\begin{split}
			&\tr_{\mathcal{G}}\qty(V\Gamma V^\dagger) =%
			\sum_{k=1}^m I \otimes \bra{u_k} \qty(\sum_{j=1}^m\sum_{i=1}^m  E_j\Gamma E_i^\dagger \otimes \ketbra{u_j}{u_i}) I\otimes\ket{u_k}\\
			&= \sum_{k=1}^m\sum_{j=1}^m\sum_{i=1}^m E_j\Gamma E_i^\dagger \underbrace{\braket{u_k}{u_j}}_{=\delta_{kj}}\underbrace{\braket{u_i}{u_k}}_{= \delta_{ik}}\\
			&= \sum_{k=1}^m E_k \Gamma E_k^\dagger\\
			&= \mathcal{E}\qty[\Gamma].
		\end{split}
		\]
		Now, we verify that $V$ is an isometry -- using Eq.~\ref{choi-kraus_isometry} of Theorem \ref{teo_kraus} to get the last equality:
		\[
		\begin{split}
			V^\dagger V &= \qty(\sum_{k=1}^m E_k^\dagger\otimes \bra{u_k}) \qty(\sum_{j=1}^m E_j \otimes \ket{u_j})%
			=  \sum_{k,j=1}^m E_k^\dagger E_j  \otimes \underbrace{\braket{u_k}{u_j}}_{= \delta_{k j}}%
			=\sum_{k=1}^m E_k^\dagger E_k%
			= I,
		\end{split}
		\]
		Finally, since $VV^\dagger$ is evidently Hermitian, to prove that it is an orthogonal projection, it suffices to verify that it is also idempotent \cite{Conway2007}. Using
		\[
		\begin{split}
			VV^\dagger &= \qty(\sum_{k=1}^m E_k \otimes \ket{u_k})\qty(\sum_{j=1}^m E_j^\dagger \otimes \bra{u_j})%
			=  \sum_{j,k=1}^m E_k E_j^\dagger \otimes \ketbra{u_k}{u_j},
		\end{split}
		\]
		we finally get -- again from Theorem \ref{teo_kraus}:
		\[
		\begin{split}
			\qty(VV^\dagger)\qty(VV^\dagger) &= \qty(\sum_{j,k=1}^m E_k E_j^\dagger \otimes \ketbra{u_k}{u_j})%
			\qty(\sum_{\ell,i=1}^m E_\ell E_i^\dagger \otimes \ketbra{u_\ell}{u_i})\\
			& = \sum_{j,k,\ell,i=1}^m E_k E_j^\dagger E_\ell E_i^\dagger \otimes \ket{u_k}\underbrace{\braket{u_j}{u_\ell}}_{=\delta_{j\ell}}\bra{u_i}\\
			&= \sum_{k,i=1}^m E_k \underbrace{\qty(\sum_{j=1}^{m}E_j^\dagger E_j)}_{=I_\mathcal{H}} E_i^\dagger \otimes \ket{u_k}\bra{u_i}\\
			&= \sum_{k,i=1}^m E_k E_i^\dagger \otimes \ket{u_k}\bra{u_i}\\
			&= VV^\dagger.
		\end{split}
		\]
		\cqd
	\end{proof}
\end{theorem}

From this point on, our discussion will be limited to quantum operations from a Hilbert $\mathcal{H}$ space to itself. In this case, the isometry property of the Stinespring dilation $V$ allows us to extend it to a unitary operator on the extended Hilbert space $\mathcal{H} \otimes \mathcal{G}$.

\begin{definition}[unitary dilation]\label{def_unitario-stinespring}
	Let $\mathcal{E}: \mathcal{L}\qty(\mathcal{H}) \rightarrow \mathcal{L}\qty(\mathcal{H})$ be a quantum channel on a Hilbert space $\mathcal{H}$ and $V: \mathcal{H} \rightarrow \mathcal{H} \otimes \mathcal{G}$ be a Stinespring dilation of $\mathcal{E}$ with ancilla space $\mathcal{G}$. Given a normalized state $\ket{u_0} \in \mathcal{G}$, a unitary dilation corresponding to the pair $\qty(V,\ket{u_0})$ is an unitary operator
	\[
	W: \mathcal{H}\otimes \mathcal{G} \rightarrow \mathcal{H}\otimes \mathcal{G}
	\]
	such that
	\[
	W\qty(\ket{\psi} \otimes \ket{u_0}) = V\ket{\psi},\ \forall \ket{\psi} \in \mathcal{H}. 
	\]
	We call the state $\ket{u_0}$ the \emph{primary ancilla state} of the unitary dilation $W$.
\end{definition}

\begin{theorem}[Environmental Representation]\label{teo_unitario-stinespring_acao}
	Under the conditions of Definition \ref{def_unitario-stinespring}, the unitary dilation $W$ corresponding to the pair $\qty(V,\ket{u_0})$ satisfies
	\begin{equation}\label{eq_waction}
		\mathcal{E}\qty[\Gamma] = \tr_\mathcal{G}\qty[W \qty(\Gamma \otimes \ketbra{u_0})W^\dagger],\ \forall \Gamma \in \mathcal{L}\qty(\mathcal{H}).
	\end{equation}
\end{theorem}
\begin{proof} 
	Let $\qty{\ket{\psi_\ell};\ \ell = 1, \dots N}$ be an orthonormal basis on $\mathcal{H}$ and take an operator $\Gamma \in \mathcal{L}\qty(\mathcal{H})$. We can write, 
	\[
	\Gamma = \sum_{i,j=1}^N \gamma_{ij}\ketbra{\psi_i}{\psi_j}\quad \qty(\gamma_{ij} \in \mathbb{C},\ \forall i,j=1. \dots, N).
	\]
	Therefore:
	\[
	\begin{split}
		\tr_\mathcal{G}\qty(W \qty(\Gamma \otimes \ketbra{u_0})W^\dagger) &=%
		\sum_{i,j=1}^N \gamma_{ij}\tr_\mathcal{G}\qty(W \qty(\ketbra{\psi_i}{\psi_j} \otimes \ketbra{u_0})W^\dagger)\\	&= \sum_{i,j=1}^N \gamma_{ij}\tr_\mathcal{G}\qty(\qty(W\ket{\psi_i \otimes u_0})\qty(W\ket{\psi_j \otimes u_0})^\dagger)\\
		&= \sum_{i,j=1}^N \gamma_{ij}\tr_\mathcal{G}\qty(\qty(V\ket{\psi_i })\qty(V\ket{\psi_j})^\dagger)\\
		&= \sum_{i,j=1}^N \gamma_{ij}\tr_\mathcal{G}\qty(V\ket{\psi_i }\bra{\psi_j}V^\dagger)\\
		&= \tr_\mathcal{G}\qty(V\Gamma V^\dagger)\\
		&=\mathcal{E}\qty(\Gamma).
	\end{split}
	\]
	\cqd
\end{proof}

\begin{theorem}[Existence of a unitary dilation]\label{teo_unitario-stinespring_exist}
		Let $\mathcal{E}: \mathcal{L}\qty(\mathcal{H}) \rightarrow \mathcal{L}\qty(\mathcal{H})$ be a quantum channel on a Hilbert space $\mathcal{H}$, $V: \mathcal{H} \rightarrow \mathcal{H} \otimes \mathcal{G}$ be a Stinespring dilation of $\mathcal{E}$ with ancilla space $\mathcal{G}$, and $\ket{u_0}\in \mathcal{G}$ a unitary statevector. Then, there exists a unitary dilation $W:\mathcal{H}\otimes \mathcal{G}\rightarrow \mathcal{H}\otimes \mathcal{G}$ corresponding to the pair $\qty(V,\ket{u_0})$.
		
	\begin{proof}
		Let $N = \dim\qty(\mathcal{H})$, $\qty{\ket{\psi_\ell};\ \ell = 1, \dots N}$ be an orthonormal basis of $\mathcal{H}$, and $\qty{\ket{u_k};\ k = 1, \dots m}$ be an orthonormal basis of $\mathcal{G}$ with $\ket{u_1} = \ket{u_0}$. We define the following orthonormal basis for the image $V(\mathcal{H}) \subset \mathcal{H}\otimes \mathcal{G}$: 
		\[ 
		\ket{\phi_{\ell,1}}:= V\ket{\psi_\ell} \in \mathcal{H}\otimes \mathcal{G},\ \forall \ell = 1, \dots N. 
		\] 
		The orthonormality of this basis follows directly from the isometry of $V$: 
		\[
		\begin{split}
			\braket{\phi_{\ell,1}}{\phi_{\tau,1}} &= \matrixel{\psi_\ell}{V^\dagger V}{\psi_\tau} %
			= \braket{\psi_\ell}{\psi_\tau}%
			= \delta_{\ell \tau},\ \forall \ell, \tau = 1, \dots, N.
		\end{split}
		\]
		Choose an orthonormal basis for the orthogonal complement $V(\mathcal{H})^\perp$, whose dimension is $N(m-1)$: 
		\[ 
		V(\mathcal{H})^\perp = \mathrm{span}\qty{\ket{\phi_{\ell,k}};\ \ell=1, \dots, N;\ k = 2, \dots, m}. 
		\] 
		Thus, we have two orthonormal bases for $\mathcal{H}\otimes \mathcal{G}$; consequently, we can define a unitary operator $W:\mathcal{H}\otimes \mathcal{G} \rightarrow \mathcal{H}\otimes \mathcal{G}$ that maps one to the other and extend it to the entire space via linearity:
		\begin{equation}\label{eq_unitario_stinespring} 
			\begin{split}
				W(\ket{\psi_\ell\otimes u_j}) := \ket{\phi_{\ell,j}},\ \forall \ell= 1, \dots, N;\ \forall j=1, \dots,m. 
			\end{split}
		\end{equation} 
		Finally, we verify that $W$ is a unitary dilation corresponding to the pair $\qty(V,\ket{u_0})$ by direct inspection: 
		\[ 
		\begin{split}
			W\qty(\ket{\psi}\otimes \ket{u_0}) &= W\qty(\ket{\psi}\otimes \ket{u_1})%
			= \ket{\phi_{\ell,1}}%
			= V\ket{\psi},\ \forall \ket{\psi} \in \mathcal{H}. 
		\end{split}
		\] 
		\cqd
	\end{proof}
\end{theorem}
\begin{remark}
	Eventually, we say that $W$ is the unitary dilation for the quantum channel $\mathcal{E}$ with primary ancilla state $\ket{u_0}$ without mention the Stinespring dilation $V$.
\end{remark}

\begin{remark}
	For developments related to Theorem \ref{teo_unitario-stinespring_acao} concerning the \emph{environmental representation}, see \cite{Bengtsson2017}, or \cite{Wolf2012}, which refer to this as an \emph{open-system representation}.
\end{remark}

\subsection{UD strategy for quantum simulation of quantum channels}\label{section_sdstrategy} 

With the goal of simulating quantum channels on digital quantum computers, we present the following proposition.

\begin{proposition}\label{pro_expval}
	Let $\mathcal{E}: \mathcal{H}\rightarrow \mathcal{H}$ be a quantum channel and $W: \mathcal{H}\otimes \mathcal{G} \rightarrow \mathcal{H}\otimes \mathcal{G}$ a unitary dilation with primary ancilla state $\ket{u_0} \in \mathcal{G}$. Given $\ket{\psi} ,\ket{\phi} \in \mathcal{H}$, it holds:
	\begin{equation}\label{eq_coq_matrixel}
		\expval{\mathcal{E}\qty(\ketbra{\psi})}{\phi} = \sum_{k=1}^m \abs{\matrixel{\phi \otimes u_k}{W}{\psi \otimes u_0}}^2,
	\end{equation}
	where $\qty{\ket{u_k};\ k = 1, \dots, m}$ is an orthonormal basis of $\mathcal{G}$.
	
	\begin{proof}
		Direct verification:
		\[
		\begin{split}
			&\expval{\mathcal{E}\qty(\ketbra{\psi})}{\phi} = \expval{\tr_\mathcal{G}\qty(W\qty(\ketbra{\psi}\otimes\ketbra{u_0})W^\dagger)}{\phi}\\
			&\quad = \sum_{k=1}^m\qty(\bra{\phi}\otimes I_\mathcal{G})\qty(I_\mathcal{H}\otimes \bra{u_k})\qty[W\qty(\ket{\psi}\otimes \ket{u_0})]%
			\qty[W\qty(\ket{\psi}\otimes \ket{u_0})]^\dagger\qty(I_\mathcal{H}\otimes \ket{u_k})\qty(\ket{\phi}\otimes I_\mathcal{G})\\
			&\quad = \sum_{k=1}^m
			\qty[\qty(\bra{\phi}\otimes \bra{u_k})W\qty(\ket{\psi}\otimes \ket{u_0})]%
			\times \qty[\qty(\bra{\phi}\otimes \bra{u_k})W\qty(\ket{\psi}\otimes \ket{u_0})]^\dagger\\
			&\quad =\sum_{k=1}^m \abs{\matrixel{\phi \otimes u_k}{W}{\psi \otimes u_0}}^2.
		\end{split}
		\]
		\cqd
	\end{proof}
\end{proposition}

The construction of quantum circuits to simulate a quantum channel $\mathcal{E}$ on a given pure state $\rho = \ketbra{\psi}$ via Eq.~\ref{eq_waction} follows these steps:
\begin{enumerate}
	\item Define sets of qubits to represent the Hilbert space $\mathcal{H}$ and the ancilla space $\mathcal{G}$, say $q$ and $a$, respectively;
	\item Assuming that all qubits are initialized in the state $\ket{\mathbf{0}}$ by default, apply gates $\mathrm{U_\psi}$ and $\mathrm{U_0}$ performing the action of unitaries $U_\psi: \mathcal{H} \rightarrow \mathcal{H}$ and $U_0: \mathcal{G} \rightarrow \mathcal{G}$ to prepare the states $\ket{\psi}$ and $\ket{u_0}$:
	\[
	U_\psi \ket{\mathbf{0}} = \ket{\psi},\ U_0 \ket{\mathbf{0}} = \ket{u_0}.
	\]
	\item Apply the unitary dilation $W$ to both qubits $\mathrm{q}$ and $\mathrm{a}$.
	\end{enumerate}
	
	Now we have to alternatives to go through:
	\begin{enumerate}
	\item[4a.] To generate or prepare the state $\mathcal{E}\qty(\ketbra{\psi})$: measure the ancilla qubits.\\
	\item[4b.] To get the matrix elements $\abs{\matrixel{\phi \otimes u_k}{W}{\psi \otimes u_0}}^2$ and the expectation value $\expval{\mathcal{E}\qty(\ketbra{\psi})}{\phi}$ in view of Eq.~\ref{eq_coq_matrixel}:
	\begin{enumerate}
		\item[4b.1.] before measure the ancilla qubits, apply the unitary $U_\phi: \mathcal{H}\rightarrow \mathcal{H}$ which prepares the state $\ket{\phi}$ from the default state $\ket{0}$,
		\[
			U_\phi \ket{\mathbf{0}} = \ket{\phi}.
		\]
		\item[4b.2] Measure the system qubits.
	\end{enumerate}
\end{enumerate}

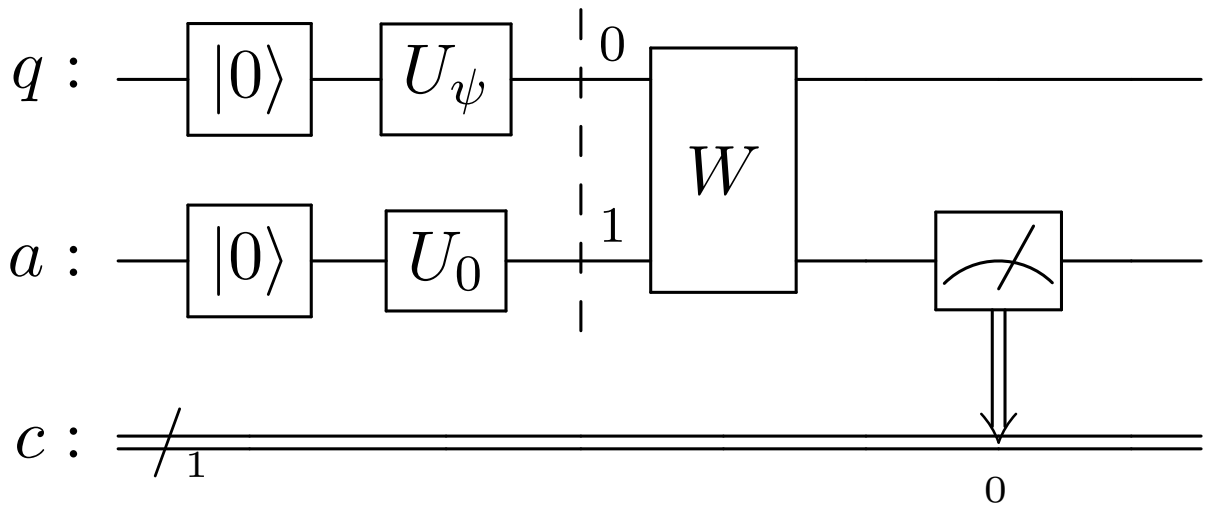
\begin{figure}
	\scalebox{1.2}{
		\Qcircuit @C=1.0em @R=1.0em @!R { \\
			\nghost{{q} :  } & \lstick{{q} :  } & \gate{ \left|0\right\rangle } & \gate{ U_\psi } \barrier[0em]{1} & \qw & \multigate{1}{ W }_<<<{0} & \qw & \qw & \qw & \qw\\
			\nghost{{a} :  } & \lstick{{a} :  } & \gate{ \left|0\right\rangle } & \gate{ U_0 } & \qw & \ghost{ W }_<<<{1} & \qw & \meter & \qw & \qw\\
			\nghost{ {c} :   } & \lstick{ {c} :   } & \lstick{/_{_{1}}} \cw & \cw & \cw & \cw & \cw & \dstick{_{_{\hspace{0.0em}0}}} \cw \ar @{<=} [-1,0] & \cw & \cw\\
			\\ }}
	\vspace{-5mm}
	\caption{\label{fig_channel}Quantum circuit to simulate a quantum channel $\mathcal{E}$ applied to the state pure $\rho = \ketbra{\psi}$, according with Theorem \ref{teo_unitario-stinespring_acao}. $q$ represents the set of system qubits, $a$ the set of auxiliary qubits, and $c$ the classical registers.}
\end{figure}

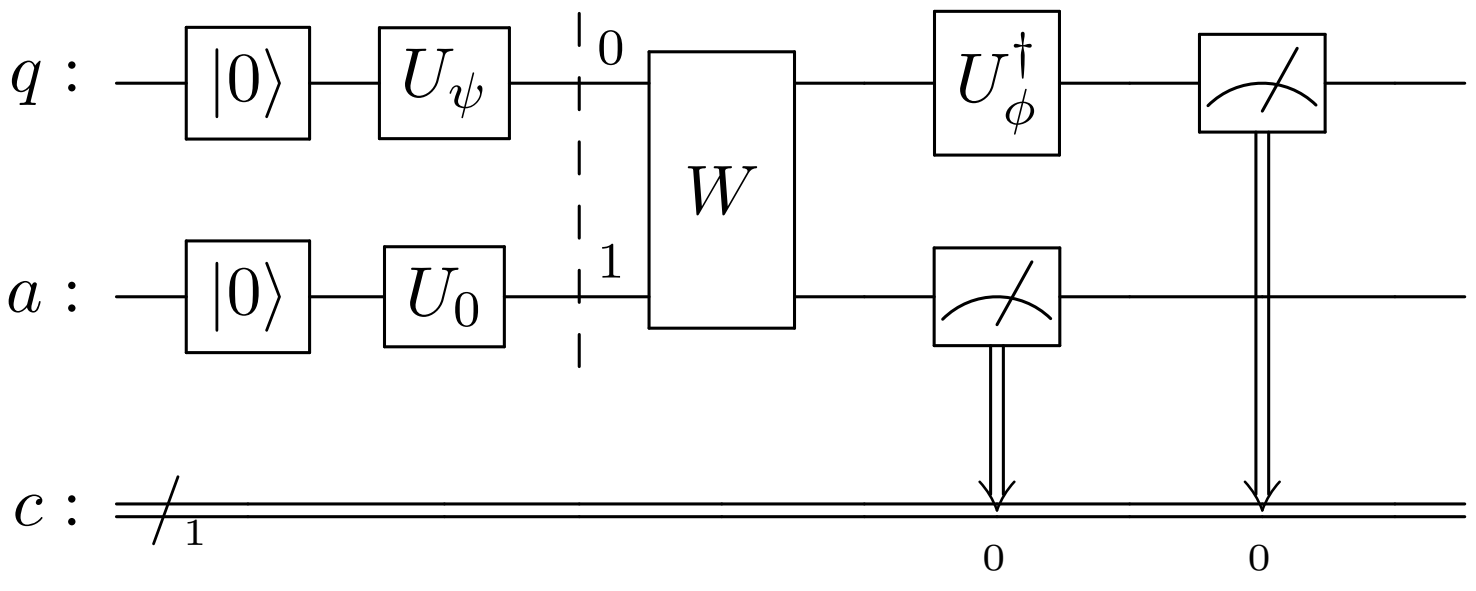
\begin{figure}
	\scalebox{1.2}{
		\Qcircuit @C=1.0em @R=1.0em @!R { \\
			\nghost{{q} :  } & \lstick{{q} :  } & \gate{ \left|0\right\rangle } & \gate{ U_\psi } \barrier[0em]{1} & \qw & \multigate{1}{ W }_<<<{0} & \qw & \gate{ U_\phi^\dagger }  & \qw & \meter & \qw & \qw\\
			\nghost{{a} :  } & \lstick{{a} :  } & \gate{ \left|0\right\rangle } & \gate{ U_0 } & \qw & \ghost{ W }_<<<{1} & \qw & \meter & \qw & \qw & \qw & \qw\\
			\nghost{ {c} :   } & \lstick{ {c} :  } & \lstick{/_{_{1}}} \cw & \cw & \cw & \cw & \cw & \dstick{_{_{\hspace{0.0em}0}}} \cw \ar @{<=} [-1,0] & \cw &  \dstick{_{_{\hspace{0.0em}0}}} \cw \ar @{<=} [-2,0] & \cw & \cw\\
			\\ }}
	\vspace{-5mm}
	\caption{\label{fig_channel-expval}Quantum circuit to evaluate the expectation value $\expval{\mathcal{E}\qty(\ketbra{\psi})}{\phi}$, according with Proposition \ref{pro_expval}. $q$ represents the set of system qubits, $a$ the set of auxiliary qubits, and $c$ the classical registers.}
\end{figure}

Figure \ref{fig_channel} shows the general form of the quantum circuit to generate or prepare the state $\mathcal{E}\qty(\ketbra{\psi})$, while Figure \ref{fig_channel-expval} shows the general form of the quantum circuit to estimate the matrix elements $\abs{\matrixel{\phi \otimes u_k}{W}{\psi \otimes u_0}}^2$ and also the expectation value $\expval{\mathcal{E}\qty(\ketbra{\psi})}{\phi}$.

Below, we discuss how to perform the quantum computational simulation of a composition of quantum channels. This is fundamental for the simulation of sequences of measurement operators, which includes as special case the quantum simulation of the evolution of Markovian quantum systems.

\subsection{Composition of quantum channels}\label{section_composition}

The composition of quantum channels remains a quantum channel, and we can obtain the Kraus operators of the composition directly from the Kraus operators of the constituent quantum channels. Let us consider the case of the composition of two quantum channels, which is sufficient to understand the general case 

Consider two quantum channels $\mathcal{E}_1: \mathcal{L}\qty(\mathcal{H}) \rightarrow \mathcal{L}\qty(\mathcal{H}')$ and $\mathcal{E}_2: \mathcal{L}\qty(\mathcal{H}') \rightarrow \mathcal{L}\qty(\mathcal{H}'')$, defined by Choi-Kraus representations of orders $m_1$ and $m_2$, respectively:
\[
\mathcal{E}_1\qty(\Gamma) = \sum_{k=1}^{m_1} E_k \Gamma E_k^\dagger;\ \ \mathcal{E}_2\qty(\Gamma') = \sum_{j=1}^{m_2} F_j \Gamma' F_j^\dagger.
\]
Then, the Choi-Kraus representation of $\mathcal{E}_2 \circ \mathcal{E}_1: \mathcal{L}\qty(\mathcal{H}) \rightarrow \mathcal{L}\qty(\mathcal{H}'')$ is given by
\begin{equation}\label{eq_coq}
	\mathcal{E}_2\circ \mathcal{E}_1\qty(\Gamma) = \sum_{k=1}^{m_1}\sum_{j=1}^{m_2} F_jE_k \Gamma E_k^\dagger F_j^\dagger,\ \forall \Gamma \in \mathcal{L}\qty(\mathcal{H}).
\end{equation}
The condition for trace preservation can be verified by direct inspection:
\[
\begin{split}
	\sum_{k=1}^{m_1}\sum_{j=1}^{m_2} E_k^\dagger F_j^\dagger F_jE_k &=\sum_{k=1}^{m_1} E_k^\dagger \qty(\sum_{j=1}^{m_2}F_j^\dagger F_j)E_k%
	= \sum_{k=1}^{m_1} E_k^\dagger E_k%
	= I.
\end{split}
\]
Thus, we obtain a Choi-Kraus representation for the composition of quantum channels $\mathcal{E}_2 \circ \mathcal{E}_1$ with an order equal to the product of the orders of the representations involved, \textit{viz.}, $m_1m_2$. For the sake of completeness, we derive in the next subsection an explicit expression for a unitary dilation for this composition 

Simulating a sequence of composite quantum operations on a quantum computer based on Eq.~\ref{eq_coq} is inefficient at least in the sense that the dimension of the ancilla Hilbert space (given by the product of the orders of the Choi-Kraus representations) grows exponentially with the number of quantum operations in the composition. However, is we can recicle ancilla qubits, then the dimension of the ancilla space depends only on the maximum order of the Choi-Kraus representations involved 

To see this, let $W_s : \mathcal{H}\otimes \mathcal{G}_s\rightarrow \mathcal{H}\otimes\mathcal{G}_s$ be the unitary dilation of the quantum channel $\mathcal{E}_s$, $\ket{u^s_0}\in \mathcal{G}_s$ be its primary ancilla state, and $\tr_s$ the partial trace over the space $\mathcal{G}_s$, for $s = 1, \dots, M$. Then, we can express the composition of the quantum operations directly in terms of their dilated unitaries:
\begin{equation}\label{eq_coa}
	\begin{split}
		\mathcal{E}_2 \circ \mathcal{E}_1 \qty[\Gamma] &= \tr_{2}\hspace{-1mm}\left( W_2\left(\tr_1\qty(W_1\qty(\Gamma \otimes \ketbra{u^1_0})W_1^\dagger)\right.\right.\hspace{-2mm}\otimes
		\left. \left.\ketbra{u^2_0}\right)W_2^\dagger\right),\ \forall \Gamma \in \mathcal{L}\qty(\mathcal{H}).
	\end{split}
\end{equation}
By induction, we have the following expression for the composition of $M$ quantum channels:
\begin{equation}\label{eq_coq_q}
	\begin{split}
		\mathcal{E}_M \circ \dots \circ \mathcal{E}_1 \qty[\rho]%
		&=\tr_{M}\left( W_M \left(  \dots \tr_1\qty(W_1\qty(\rho \otimes \ketbra{u^1_0})W_1^\dagger)\otimes\right.\right.\dots\\ %
		&\qquad \qquad \qquad \dots \left.\left.\otimes \ketbra{u^M}\right)W_M^\dagger\right),\ \forall \rho \in \mathcal{L}\qty(\mathcal{H}).
	\end{split}
\end{equation}

\begin{figure}
	\scalebox{1.0}{
		\Qcircuit @C=1.0em @R=1.0em @!R { \\
			\nghost{{q} : } & \lstick{{q} : } &\gate{\left|0\right\rangle} &\gate{U_\psi} 
			\barrier[0em]{1} &\qw 
			&\multigate{1}{W_1}_<<<{0} &\qw &\qw  &\qw &\qw
			&\dots &
			\barrier[0em]{1} &\qw
			&\multigate{1}{W_M}_<<<{0} &\qw &\qw &\qw
			& \qw & \qw\\
			\nghost{{a} :  } & \lstick{{a} :  } &\gate{\left|0\right\rangle} &\qw &\qw
			&\ghost{W}_<<<{1} &\meter &\gate{\left|0\right\rangle} &\qw &\qw
			&\dots &
			&\qw  &\ghost{W}_<<<{1}  &\qw & \meter &\gate{\left|0\right\rangle} &\qw &\qw\\
			\nghost{{c} :  } & \lstick{{c} :  } & \lstick{/_{_{1}}} \cw &\cw &\cw &\cw &\cw
			&\dstick{_{_{\hspace{0.0em}0}}} \cw \ar @{<=} [-1,0] &\cw &\cw & \dots & &\cw &\cw &\cw &\dstick{_{_{\hspace{0.0em}0}}} \cw \ar @{<=} [-1,0] &\cw &\cw &\cw\\
			\\ }}
	\caption{\label{fig_channel-composition}Quantum circuit that simulates the action of a composition of quantum channels on a (pure) state $\rho = \ketbra{\psi}$, with prime ancillary states being $\ket{u_0^j} = \ket{0}$ ($j = 1, \dots , q$)}.
\end{figure}
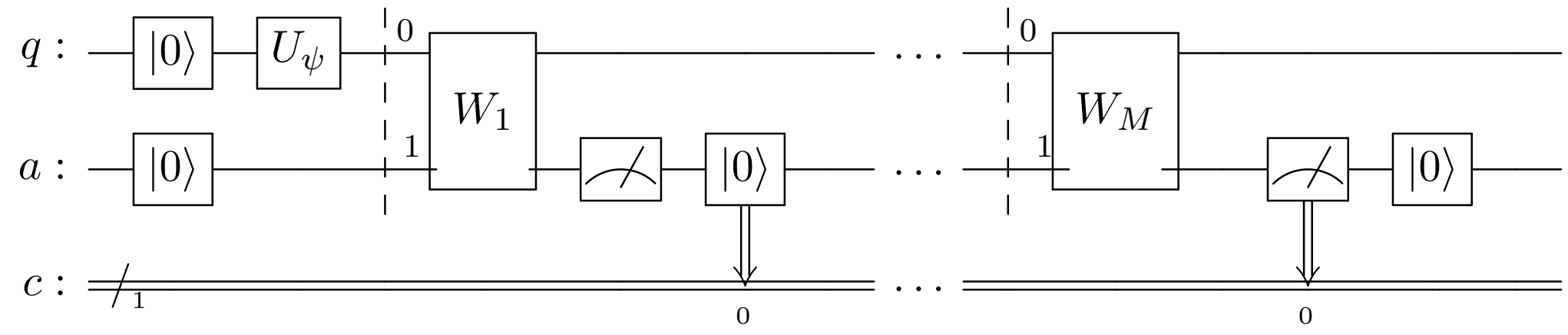

Figure \ref{fig_channel-composition} shows the quantum circuit for simulation of the action of this composition of quantum channels on a (pure) state $\rho = \ketbra{\psi}$, with prime ancillary states being $\ket{u_0^j} = \ket{0}$ ($j = 1, \dots , q$). When the state $\rho \in \mathcal{D}\qty(\mathcal{L})$ is mixed, the gate $U_\psi$ in this figure can be replaced by the procedure from Section \ref{secao_prepest} for preparing it.

\subsubsection{Unitary dilation for a composition of quantum channels}

The unitary dilation associated to the Choi-Kraus representation for the composition of two quantum channels given by Eq.~\ref{eq_coq} can be easily derived using \emph{trace duality}, which allows us to commute a unitary dilation with the partial trace on Hilbert spaces where the former acts trivially:

\begin{lemma}[Trace duality]\label{lema_dualidade-traco}
	Let $\mathcal{H}_A$ and $\mathcal{H}_B$ be Hilbert spaces. Given $A \in \mathcal{L}(\mathcal{H}_A)$ and $C \in \mathcal{L}(\mathcal{H}_A\otimes \mathcal{H}_B)$, the following holds:
	\[
	\tr_A\qty(A\tr_B\qty(C)) = \tr_{AB}\qty(\qty(A \otimes I_B)C). 
	\]
	For $A_1, A_2\in \mathcal{L}\qty(\mathcal{H}_A)$ and $C \in \mathcal{L}(\mathcal{H}_A\otimes \mathcal{H}_B)$, the following also holds:
	\[
	\tr_A\qty(A_1\tr_B\qty(C) A_2) = \tr_{AB}\qty(\qty(A_1 \otimes I_B) C \qty(A_2 \otimes I_B)). 
	\]
\end{lemma}
The proof of this lemma follows directly from the definition of the partial trace -- see, for example \cite[Example 4.2, p.181]{Heinosaari2011}. In particular, the second identity is a direct consequence of the first combined with the cyclicity of the trace:
\[
\begin{split}
	\tr_A\qty(A_1 \tr_B\qty(C)A_2) &=  \tr_A\qty(A_2A_1 \tr_B\qty(C))\\
	&= \tr_{AB}\qty(\qty(A_2 \otimes I_B)\qty(A_1 \otimes I_B)C)\\
	&= \tr_{AB}\qty(\qty(A_1 \otimes I_B)C\qty(A_2 \otimes I_B)). 
\end{split}
\]

Using trace duality, we can commute the unitary dilation of one quantum channel with the partial trace in the ancilla space of another. To this end, we extend the dilated unitaries of each quantum channel to the tensor product $\mathcal{G}_1\otimes\mathcal{H}\otimes  \mathcal{G}_2$ using the identities $I_s \in \mathcal{G}_s$, for $s = 1, 2$:\footnote{
	We have placed the Hilbert space $\mathcal{H}$ in the middle of the tensor product merely to simplify the notation.}
\[
\tilde{W}_1 := W_1 \otimes I_2\ \ \text{and}\ \ \tilde{W}_2 :=  I_1 \otimes W_2.
\]
Explicitly,
\[
\begin{split}
	&\left\lbrace
	\begin{array}{l}
		\tilde{W}_1\qty(\ket{\phi} \otimes \ket{\psi}\otimes \ket{\varphi}) = W_1\qty( \ket{\phi}\otimes\ket{\psi}) \otimes \ket{\varphi},\\
		\tilde{W}_2\qty(\ket{\phi} \otimes \ket{\psi}\otimes \ket{\varphi}) =  \ket{\phi}\otimes W_2\qty(\ket{\psi} \otimes \ket{\varphi});
	\end{array}
	\right.\\
	& \quad \quad \quad \quad \quad \quad \quad \forall \ket{\psi} \in \mathcal{H}, \ket{\phi} \in \mathcal{G}_1, \ket{\varphi} \in \mathcal{G}_2.
\end{split}
\]
Now, using trace duality on Eq.~\ref{eq_coq_q} we get
\begin{equation}\label{eq_coq_2}
	\begin{split}
		\mathcal{E}_2 \circ \mathcal{E}_1 \qty[\rho]%
		&= \tr\qty[\qty(I_1 \otimes W_2)\qty(W_1\otimes I_2)
		 \qty(\ketbra{u^1_0} \otimes \rho \otimes \ketbra{u^2_0})%
		 \qty(W_1\otimes I_2)^\dagger\qty(I_1 \otimes W_2)^\dagger]\\
		&=\tr\hspace{-1mm}\qty(\tilde{W}_2\tilde{W}_1\qty(\ketbra{u^1_0}\hspace{-1mm} \otimes\hspace{-1mm} \rho \hspace{-0.5mm} \otimes\hspace{-1mm} \ketbra{u^2_0})\tilde{W}_1^\dagger\tilde{W}_2^\dagger )%
		,\ \forall \rho \in \mathcal{L}\qty(\mathcal{H}). 
	\end{split}
\end{equation}

In summary, $\tilde{W}_2 \tilde{W}_1$ in Eq.~\ref{eq_coq_2} is a unitary dilation for the Choi-Kraus representation defined in Eq.~\ref{eq_coq}, with an ancilla state space $\mathcal{G}_2 \otimes \mathcal{G}_1$ and an associated primary ancilla state $\ket{u_0^2}\otimes \ket{u_0^1}$.

\subsection{Computational complexity}

In this section, we analyze the computational resources required for the quantum computational simulation of a composition of $M \ge 1$ quantum channels acting on a quantum system with a Hilbert space $\mathcal{H}$ of dimension $N$, based on Eq.~\ref{eq_coq_q} and Eq.~\ref{eq_coq_2}. Obviously, the computational complexity for the quantum simulation of a single quantum channel is obtained by setting $M=1$ 

For each quantum channel $\mathcal{E}_s$ in the composition ($s = 1, \dots, M$), we denote $\mathcal{G}_s$ as the ancilla state space, $m_s = \dim(\mathcal{G}_s)$ as the dimension of the Choi-Kraus representation, and $\tilde{W}_s$ as the associated unitary dilation.

\textbf{Number of qubits.} To simulate the composition of $M$ quantum channels, we require $n = \lceil \log_2 N \rceil$ qubits to represent the physical system, plus $\lceil \log_2 m_s \rceil$ ancilla qubits to represent each ancilla state space, resulting in a total number of qubits given by
\[
\log_2 \qty(Nm_1\dots m_M).
\]
The number of ancilla qubits can be significantly reduced by recycling them, provided the quantum computer is capable of such recycling -- that is, to perform partial measurements and reinitialize qubits during the quantum processing. In this case, the number of ancilla qubits is determined by the highest order of the Choi-Kraus representations:
\[
\log_2 N + \log_2 \tilde{m},\ \tilde{m}: = \max\qty{m_1, \dots, m_M}.
\]
\textbf{Number of two-qubit gates.} A gate of dimension $D \times D$ ($D \ge 2$) can be obtained as a composition of up to $D(D-1)/2$ two-qubit gates \cite{Nielsen2010}. Thus, assuming the initial state of the physical system simulation is a pure state, preparing it requires up to $N(N-1)/2$ two-qubit gates; similarly, the unitary dilation $W_s$ has dimension $N m_s$, so its implementation requires up to $Nm_s(Nm_s -1)/2$ two-qubit gates;%
\footnote{Since $\tilde{W}_s$ is the tensor product of $W_s$ with an identity operator, the decomposition of both requires the same number of two-qubit gates. 
} %
therefore, the number of two-qubit gates required to simulate the composition of quantum channels is
\[
\frac{N(N-1)}{2} + \sum_{s = 1}^M \frac{Nm_s(Nm_s -1)}{2}.
\]
When the orders of all Choi-Kraus representations are the same ($m_s = m, \forall s = 1, \dots, M$), we have
\begin{equation}\label{}
\begin{split}\label{cc_totalqubits}
	&\frac{N}{2}\left[N - 1 + M m(Nm -1)\right] \sim O\left(MN^2m^2\right).
\end{split}
\end{equation}
This number of two-qubit gates does not decrease with the reuse of ancilla qubits; on the contrary, such reuse may increase the number of two-qubit gates, to the extent necessary to reset the ancilla qubits.

\section{\label{secao_prepest}Preparation of mixed states}

In this section, we show how to prepare mixed states using quantum channels constructed from their spectral decomposition. By combining this with the strategy for simulating quantum channels presented in Section \ref{secao_canaisq}, we obtain a protocol for preparing mixed states on digital quantum computers 

Let $\rho \in \mathcal{D}\qty(\mathcal{H})$ be a generic state (density operator) in a Hilbert space $\mathcal{H}$. Consider its spectral decomposition:
\[
\begin{split}
	\rho &= \sum_{\ell=1}^N c_\ell \ketbra{\psi_\ell};\ %
    c_\ell \ge 0, \forall \ell = 1, \dots, N;\ \sum_{\ell = 1}^N c_\ell = 1,
\end{split}
\]
where $\{\ket{\psi_\ell} : \ell = 1, \dots, N\}$ is an orthonormal basis of $\mathcal{H}$ (eigenstates of $\rho$) 

Let $\mathcal{G} := \mathbb{C}^N$ and $\{\ket{u_k} : k = 1, \dots, N\}$ be an orthonormal basis for $\mathcal{G}$. Also, let $\ket{\psi_0} \in \mathcal{H}$ be an initialization state%
\footnote{
	Typically, the qubits in digital quantum computers are initialized to the computational basis state $\ket{0}$.
}%
, and choose unitary operators $U_\ell: \mathcal{H} \rightarrow \mathcal{H}$ such that
\[
U_\ell \ket{\psi_0} = \ket{\psi_\ell},\ \forall \ell \in \{1, \dots, N\}.
\]
We define the following quantum operation directly in the Choi-Kraus representation:
\[
\mathcal{E}_\rho \qty[\Gamma] := \sum_{\ell = 1}^N E_\ell \Gamma E_\ell^\dagger,\ \forall \Gamma \in \mathcal{L}\qty(\mathcal{H}),
\]
where
\[
E_\ell := \sqrt{c_\ell}U_\ell, \forall \ell = 1, \dots N.
\]

We can verify by direct inspection that $\mathcal{E}_\rho$ indeed prepares $\rho$ from the pure state $\ket{\psi_0}$ and satisfies the trace-preserving condition:
\begin{equation}\label{eq_prepest}
\begin{split}
	\mathcal{E}_\rho \qty[\ketbra{\psi_0}] &= \sum_{\ell = 1}^N E_\ell \ketbra{\psi_0} E_\ell^\dagger%
	= \sum_{\ell = 1}^N c_\ell U_\ell \ketbra{\psi_0} U_\ell^\dagger%
	= \sum_{\ell = 1}^N c_\ell \ketbra{\psi_\ell}%
	= \rho;
\end{split}
\end{equation}
 and
\[
\sum_{\ell = 1}^N E_\ell^\dagger E_\ell = \sum_{\ell = 1}^N  c_\ell U_\ell^\dagger  U_\ell = \sum_{\ell = 1}^N  c_\ell I = I.
\]
In accordance with Theorem \ref{teo_dilatacao-stinespring}, we have the Stinespring dilation of $\mathcal{E}_\rho$ associated with its Choi-Kraus representation:
\[
V_\rho \ket{\psi}:= \sum_{\ell=1}^N \sqrt{c_\ell}\qty(U_\ell\ket{\psi})\otimes \ket{u_\ell}.
\]
We fix the primary ancilla state as
\[
\ket{u_0}:= \sum_{k=1}^N \sqrt{c_k}\ket{u_k}
\]
and define the unitary dilation corresponding to the pair $(V, \ket{u_0})$ by its action on the basis of $\mathcal{H}\otimes \mathcal{G}$:
\[
W_\rho \qty(\ket{\psi_\ell} \otimes \ket{u_k}):= \qty(U_k\ket{\psi_\ell})\otimes \ket{u_k},\ \forall  \ell, k = 1, \dots, N.
\]
Indeed, $W_\rho$ is a linear operator due to the linearity of the unitaries $U_k$. To verify its unitarity, we simply observe that it applies the orthonormal basis $\qty{\ket{\psi_\ell} \otimes \ket{u_k};\ \ell, k = 1, \dots N}$ on an orthonormal set of $\mathcal{H}\otimes \mathcal{G}$:
\[
\begin{split} 
	\matrixel{\psi_\ell \otimes u_k}{W_\rho^\dagger W_\rho}{\psi_\tau \otimes u_j} &= \qty(\bra{\psi_\ell}U_k^\dagger\otimes \bra{u_k})\qty({U_j}\ket{\psi_\tau}\otimes \ket{u_j})\\ 
	&\quad = \matrixel{\psi_\ell}{U_k^\dagger U_j}{\psi_\tau}\underbrace{\braket{u_k}{u_j}}_{=\delta_{k, j}}\\ 
	&\quad = \delta_{k, j}\matrixel{\psi_\ell}{\underbrace{U_k^\dagger U_k}_{=I}}{\psi_\tau}\\ 
	&\quad = \delta_{k, j}\delta_{\ell, \tau}, \quad \forall \ell, \tau, k, j = 1, \dots N.
\end{split}
\]
Now, to verify that $W_\rho$ corresponds to the pair $\qty(V,\ket{u_0})$. For a state $\ket{\psi} \in \mathcal{H}$, let its decomposition in the basis of $\mathcal{H}$ be:
\[
\ket{\psi} = \sum_{\ell = 1}^N c_\ell \ket{\psi_\ell};
\]
then:
\[
\begin{split}
	W_\rho  \qty(\ket{\psi} \otimes \ket{u_0}) &=
	\sum_{\ell , k = 1}^N c_\ell \sqrt{c_k} W_\rho \qty(\ket{\psi_\ell} \otimes \ket{u_k})\\%
    &=\sum_{\ell , k = 1}^N c_\ell \sqrt{c_k} \qty(U_k\ket{\psi_\ell}) \otimes \ket{u_k}\\%
    &=\sum_{\ell , k = 1}^N c_\ell \qty(E_k\ket{\psi_\ell}) \otimes \ket{u_k}\\%
    &= \sum_{\ell = 1}^N c_\ell V_\rho\ket{\psi_\ell}\\%
    &= V_\rho \qty(\sum_{\ell = 1}^N c_\ell \ket{\psi_\ell})\\%
    &= V_\rho \ket{\psi}.
\end{split}
\]
Finally, combining Eq.~\ref{eq_prepest} and Eq.~\ref{eq_waction} (with $\Gamma = \ketbra{\psi_0}$), we get 
	\begin{equation}\label{eq_purification}
	\rho = \tr_\mathcal{G}\qty(W_\rho\ket{\psi\otimes u_0}\bra{\psi\otimes u_0}W_\rho^\dagger).
	\end{equation}
\begin{remark}
	Identity of Eq.~\ref{eq_purification} means that the statevector $W\ket{\psi \otimes u_0}\mathcal{H}\otimes \mathcal{G}$ is a \emph{purification of $\rho$} in the extended Hilbert space \cite{Heinosaari2011, Nielsen2010, Wilde2016}.
\end{remark}
Now, we will show how to use the unitary dilation $W_\rho$ to obtain the expection value of an observable with respect to $\rho$.

\subsection{Expectation value of observables}

Recall that the expected value of an observable (a Hermitian operator) $A \in \mathcal{L}(\mathcal{H})$ with respect to a state $\rho \in \mathcal{D}(\mathcal{H})$ is given by 
\[
\expval{A}_\rho = \tr(\rho A).
\]

From the expression of $\rho$ in terms of unitary dilation Eq.~\ref{eq_purification}, we get the expectation value of the observable $A$ via trace duality (Lemma \ref{lema_dualidade-traco}):
\begin{equation}\label{eq_expval}
	\begin{split}
		\expval{A}_{\rho} &= \tr_\mathcal{H}(\tr_\mathcal{G}(W_\rho \qty(\rho \otimes \ketbra{u_0}) W_\rho^\dagger) A)%
		=\tr(W_\rho \qty(\rho \otimes \ketbra{u_0}) W_\rho^\dagger(A \otimes I_\mathcal{G})).
	\end{split}
\end{equation}

Below, we present an example of a quantum computational simulation that makes explicit use of Eq.~\ref{eq_expval} to obtain the expected value of an observable with respect to a mixed state.

\subsection{Example: preparation of a mixed state on a single qubit}\label{secao_prepest_exemplo}

A single qubit is described by a two dimensional Hilbert space
for which we use the computational basis to expand it:\footnote{
	Here, all matrices for operators are defined with respect to the computational basis.}
\[
\mathcal{H} = \mathrm{span}\{\ket{0}, \ket{1}\}.
\]
Let the mixed state to be prepared be 
\[
\begin{split}
	\rho &= \frac{2}{3}\ketbra{+} + \frac{1}{3}\ketbra{-},
\end{split}
\]
where we have the orthonormal vectors
\[
\begin{split}
	&\ket{\psi_1}:= \ket{+} = \frac{\ket{0} + \ket{1}}{\sqrt{2}}\ \text{and}\% \ket{\psi_2}:= \ket{-}=\frac{\ket{0} - \ket{1}}{\sqrt{2}}.
\end{split}
\]
The observable we will evaluate the expectation valued with respect to $\rho$ via quantum simulation is
\[
\begin{split}
	A &= I + 2X - 3Y + 4Z = \begin{pmatrix}
		5 & 2 + 3i\\
		2 - 3i & -3
	\end{pmatrix}%
	;\\ %
	X &= \begin{pmatrix}
		0 & 1\\
		1 & 0
	\end{pmatrix}%
	,\ %
	Y = \begin{pmatrix}
		0 & -i\\
		i & 0
	\end{pmatrix},\ %
	Z = \begin{pmatrix}
		1 & 0\\
		0 & -1
	\end{pmatrix}.
\end{split}
\]
We fix the initial state $\ket{\psi_0} := \ket{0} \in \mathcal{H}$ from which we will prepare $\rho$ and also the unitaries that prepares $\ket{\psi_1}$ and $\ket{\psi_2}$ from $\ket{\psi_0}$:
\[
\begin{split}
	U_1 &:= H \equiv \frac{1}{\sqrt{2}}%
	\begin{pmatrix}
		1 & 1\\
		1 & -1
	\end{pmatrix}\ %
	\ \text{and}\ %
	U_2 := HX \equiv \frac{1}{\sqrt{2}}%
	\begin{pmatrix}
		1 & 1\\
		-1 & 1
	\end{pmatrix},
\end{split}
\]
where we introduced the \emph{Hadamard gate}
\[
H = \frac{X + Z}{\sqrt{2}} = \frac{1}{\sqrt{2}}\begin{pmatrix}
	1 & 1\\
	1 & 1
\end{pmatrix}.
\]
It holds, as one can easily check:
\[
U_k \ket{\psi_0} = \ket{\psi_k},\ k = 1,2.
\] 
Following previous developments, we define the Kraus operators
\[
E_1 = \sqrt{2/3}H \quad \text{and}
\quad E_2 = \sqrt{1/3}HX
\]
and the quantum channel in the Choi-Kraus representation:
\[
\begin{split}
	\mathcal{E}_\rho[\Gamma] &:= E_1 \Gamma E_1^\dagger + E_2 \Gamma E_2^\dagger%
	 = \frac{2}{3}H\Gamma H + \frac{1}{3}HX\Gamma XH.
\end{split}
\]
The ancilla Hilbert space $\mathcal{G}$ has two dimensions and we fix the orthonormal basis for it to be the computational basis:
\[
\mathcal{G} := \mathrm{span}\qty{\ket{u_0},\ket{u_1}};\ %
\ket{u_0}:= \ket{0},\ \ket{u_1}:=\ket{1}.
\] 
The Stinespring dilation associated to the Choi-Kraus representation of $\mathcal{E}_\rho$ is
\[
\begin{split}
	&V_\rho: \mathcal{H}\rightarrow \mathcal{H}\otimes \mathcal{G},\ %
	V_\rho\ket{\psi}:= \qty(E_1\ket{\psi}) \otimes \ket{0} + \qty(E_2 \ket{\psi}) \otimes \ket{1}%
	= H\ket{\psi \otimes 0} + HX\ket{\psi \otimes 1}.
\end{split}
\]\\
We fix the primary ancilla state to be
\[
\ket{u_0}:= \sqrt{\frac{2}{3}}\ket{0} + \sqrt{\frac{1}{3}}\ket{1} = R_y\qty(2\arctan(1/\sqrt{2})),
\]
where
\[
R_y(\theta) = \begin{pmatrix} 
	\cos(\frac{\theta}{2}) & -\sin(\frac{\theta}{2})\\ 
	\ \\
	\sin(\frac{\theta}{2}) & \cos(\frac{\theta}{2}) 
\end{pmatrix}.
\] 
then, the unitary dilation of $\mathcal{E}_\rho$ is given by:
\begin{equation}\label{eq_prepest_w}
	\begin{split}
		W_\rho \qty(\ket{\psi_\ell}\otimes \ket{u_k}) &= \qty(U_k\ket{\psi_\ell})\otimes \ket{k}\\%
		& = \left\{
		\begin{array}{rl}
			H\ket{\ell} \otimes \ket{0},& k = 0\ (\ell = 0,1)\\
			HX\ket{\ell} \otimes \ket{1},& k = 1\ (\ell = 0,1)
		\end{array}
		\right.\\
		& = HC_X^{0\rightarrow 1}\qty(\ket{\ell}\otimes \ket{k}),
	\end{split}
\end{equation}
where $C_X^{0\rightarrow 1}$ denotes the controlled-$X$ gate, with the qubit on the right (ancilla) acting as the control and the qubit on the left (system) acting as the target.

\begin{figure}
	\scalebox{1.0}{
		\Qcircuit @C=1.0em @R=0.2em @!R { \\
			\nghost{{q} :  } & \lstick{{q} :  } & \gate{ \left|0\right\rangle}  & \qw & \qw  & \qw & \targ & \gate{H} & \qw & \qw  & \qw & \qw & \qw\\
			\nghost{{a} :  } & \lstick{{a} :  } & \gate{\left|0\right\rangle} & \qw & \gate{R_Y\,(1.231)} & \qw & \ctrl{-1} & \qw & \qw & \meter & \qw & \qw & \qw\\
			\nghost{{c} :  } & \lstick{{c} :  } & \lstick{/_{_{1}}} \cw & \cw & \cw & \cw & \cw & \cw & \cw & \dstick{_{_{\hspace{0.0em}0}}} \cw \ar @{<=} [-1,0] & \cw & \cw & \cw\\
			\\ }}
	\vspace{-5mm}
	\caption{\label{fig_prepest_rho}Quantum circuit preparing the state $\rho$ in the qubit $\mathrm{q}$. Here, $R_y(1.231)$ corresponds to the unitary preparing the primary ancilla state $\ket{u_0}$ and the following pair of gates $C_X^{0\rightarrow 1}$ and $H$ correspond to the unitary dilation $W_\rho$ (Eq.~\ref{eq_prepest_w}).}
\end{figure}
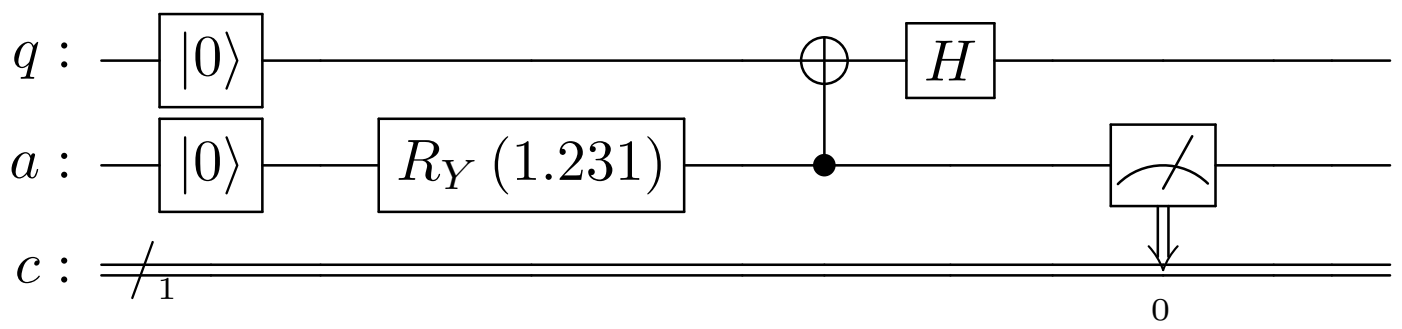

To ensure the consistency of the construction, we explicitly verify that $W_\rho$ corresponds to the pair $\qty(V,\ket{u_0})$; for that, we take an arbitrary pure state
\[
\ket{\psi} = \alpha \ket{\psi_1} + \beta\ket{\psi_2} \in \mathcal{H},
\]
and calculate the action of $W_\rho$ on it:
\[
\begin{split}
	&W_\rho \qty(\ket{\psi}\otimes \ket{u_0})  =\\%
	&=W_\rho \qty(\hspace{-1mm}\qty(\alpha \ket{\psi_1} + \beta\ket{\psi_2})\hspace{-1mm}\otimes\hspace{-1mm} \qty(\sqrt{\frac{2}{3}}\ket{u_0} + \sqrt{\frac{1}{3}}\ket{u_1})\hspace{-1mm})\\
	&= \alpha\sqrt{\frac{2}{3}} \qty( U_1\ket{\psi_1})\otimes \ket{u_0}%
	+ \alpha\sqrt{\frac{1}{3}} \qty( U_2\ket{\psi_1})\otimes \ket{u_1}
	+\\
	&\quad \quad \quad + \beta\sqrt{\frac{1}{3}} \qty(U_2 \ket{\psi_2})\otimes \ket{u_1}\\
	&= \alpha V\ket{\psi_1} + \beta V\ket{\psi_2}\\
	&= V\ket{\psi}.
\end{split}
\]
In Figure \ref{fig_prepest_rho}, we show the circuit that prepares $\rho$, with the system qubit $q$ (top) and the prime ancilla qubit $a$ (bottom).

We simulated an experiment using the circuit shown in Figure \ref{fig_prepest_rho} on the \emph{AerSimulator} from the \emph{Qiskit} SDK with $10^{6}$ shots, obtaining the following result with precision estimated of $10^{-3}$:\footnote{GitHub page for the code:\newline \url{https://github.com/LucioFassarella/SQC/blob/main/SQC_state-preparation.ipynb}.}

\[
\expval{A}_{\rho\ \text{exp}} = 1.668124.
\]
To compare, the theoretical expectation value of the observable $A$ with respect to the state $\rho$ is given by
\[
\begin{split}
	\expval{A}_\rho &= \tr\qty(\rho A) \\
	&= \underbrace{\tr(\rho I)}_{=1} + 2\underbrace{\tr(\rho X)}_{=1/3} - 3\underbrace{\tr(\rho Y)}_{=0} + 4\underbrace{\tr\qty(\rho Z)}_{=0}\\
	&= \frac{5}{3\sqrt{2}} \approx 1.666....
\end{split}
\]
As can be observed, the simulated an theoretical values for the expectation value show reasonable agreement.

\section{\label{secao_povm}Positive Operator-Valued Measures}
\begin{definition}[POVM]\label{def_povm_medida-operador}
	A \emph{positive operator-valued measure} (POVM) on a Hilbert space $\mathcal{H}$ is a finite family of operators, referred to as measurement operators,
	\[
	M = \{M_k: \mathcal{H} \rightarrow \mathcal{H}; \ k=1, \dots, m\}
	\]
	that defines a partition of identity, \textit{i.e.},
	\[
	\sum_{k=1}^m M_k^\dagger M_k = I.
	\]
	An operator-valued measure $M$ canonically defines a quantum channel in the Choi-Kraus representation, where the Kraus operators are identified with the measurement operators:
	\begin{equation}\label{eq_povmchannel}
	\mathcal{E}_M[\Gamma] := \sum_{k=1}^m M_k \Gamma M_k^\dagger, \quad \forall \Gamma \in \mathcal{L}(\mathcal{H}).
	\end{equation}
\end{definition}

In quantum mechanics, the \emph{measurement postulate} \cite{Nielsen2010, Olivares2025, Wilde2016} establishes that the measurements we can perform on quantum systems are described by operator-valued measures. 
\begin{quotation}\textbf{Measurement postulate}: The measurement of the quantum system in the state $\rho \in \mathcal{D}\qty(\mathcal{H})$ by the POVM $M = \qty{M_k;\ k = 1, m}$ obeys two conditions:
	\begin{itemize}
		\item \textbf{Born's rule}: the probability of the measurement result being $M_k$ is
		\begin{equation}\label{eq_povm_probabilidades} 
			\mathrm{Pr}\qty(M_k\vert \rho)  = \tr \qty(M_k\rho M_k^\dagger).
		\end{equation}
		\item \textbf{Post-measure state}: if the measurement result is $M_k$, then the state of the system immediately after the measurement is given by
		\begin{equation}\label{eq_povm_estado-pos}
			\rho_k = \frac{M_k \rho M_k^\dagger}{\tr \qty(M_k\rho M_k^\dagger)}.
		\end{equation}
	\end{itemize}
	We highlight two specializations:
	\begin{itemize}
		\item[$\circ$] \emph{projective measure} is a positive operator-valued measure whose measurement operators are orthogonal projections.
		\item[$\circ$] If the system is in a pure state $\rho = \ketbra{\psi}$, $\ket{\psi} \in \mathcal{H}$ (normalized), the outcome probabilities are given by
		\begin{equation}\label{eq_povm_probabilidades2} 
			\mathrm{Pr}\qty(M_k\vert \psi)  = \sum_{\ell = 1}^N \abs{\matrixel{\psi_\ell}{M_k}{\psi}}^2,
		\end{equation}
		where $\qty{\ket{\psi_\ell};\ \ell = 1, \dots N}$ is any orthonormal basis of $\mathcal{H}$.
	\end{itemize}
\end{quotation}
\begin{proposition}\label{teo_povm_unitario-stinespring}
	Under the conditions of Definition \ref{def_povm_medida-operador}, let $V_M : \mathcal{H}\rightarrow \mathcal{H}\otimes \mathcal{G}$ be a Stinespring dilation of the quantum channel $\mathcal{E}_M$, with an ancilla Hilbert space $\mathcal{G}$ equipped with an orthonormal basis $\{\ket{u_k}; k = 1, \dots, m\}$, as established by Theorem \ref{teo_dilatacao-stinespring}:
	\[
	V_M\ket{\psi} = \sum_{k=1}^m M_k\ket{\psi} \otimes \ket{u_k},\ \forall \psi \in \mathcal{H}.
	\]
	Let $W_M$ be the unitary dilation corresponding to the pair $(V_M, \ket{u_0})$, for some primary ancilla state $\ket{u_0}\in \mathcal{G}$. Then, it holds:
	\begin{equation}\label{eq_povm_stinespring_tracoH}
		\begin{split}
			\mathrm{Pr}(M_k \vert \rho) &= \matrixel{u_k}{\tr_\mathcal{H}\qty(W_M \qty(\rho \otimes \ketbra{u_0})W_M^\dagger)}{u_k},\ %
			\forall \rho \in \mathcal{D}\qty(\mathcal{H}),\ \forall k = 1, \dots, m.
		\end{split}
	\end{equation}
	In particular,
	\begin{equation}\label{eq_povm_stinespring_probabilidades}
		\begin{split}
			\mathrm{Pr}(M_k\vert \psi)  =\sum_{\ell = 1}^N \bra{\psi_\ell \otimes u_k} W_M\ket{\psi \otimes u_0},\ %
            \forall \ket{\psi} \in \mathcal{H},\ \forall k = 1, \dots, m,
		\end{split}
	\end{equation}
	where $\qty{\ket{\psi_\ell};\ \ell = 1, \dots, N}$ is any orthonormal basis of $\mathcal{H}$.
    
	\begin{proof} 
		From the definition of Stinespring dilation \ref{def_ditatacao-stinespring} and the definition of unitary dilation \ref{def_unitario-stinespring}, for any $\ket{\psi} \in \mathcal{H}$ it holds:
		\[
		\begin{split}
			W_M\qty(\ket{\psi}\otimes \ket{u_0})= \sum_{k=1}^m \qty(M_k \ket{\psi})\otimes \ket{u_k}.
		\end{split}
		\]
		From this, we get:
		\[
		\begin{split}
			\matrixel{\phi}{M_k}{\psi} &= \qty(\bra{\phi}\otimes I_\mathcal{G})\qty(I_\mathcal{H}\otimes \bra{k})W_M\qty(\ket{\psi}\otimes \ket{u_0})%
            =\bra{\phi \otimes k}W_M \ket{\psi \otimes u_0}
		\end{split}.
		\]
		Applying this identity to Eq.~\ref{eq_povm_probabilidades2}, we obtain Eq.~\ref{eq_povm_stinespring_probabilidades}:
		\[
		\begin{split}
			\mathrm{Pr}\qty(M_k\vert \psi) &= \sum_{\ell=1}^N \abs{\matrixel{\psi_\ell}{M_k}{\psi}}^2 %
            = \sum_{\ell=1}^N \abs{\matrixel{\psi_\ell \otimes k}{W_M}{\psi\otimes u_0}}^2.
		\end{split}
		\]
		To prove Eq.~\ref{eq_povm_stinespring_tracoH}, it suffices to consider the special case of pure states $\rho = \ketbra{\psi}$ ($\ket{\psi} \in \mathcal{H}$ normalized), since the general case follows from this special case via spectral decomposition. Indeed,
		\[
		\begin{split}
			&\tr_\mathcal{H}\qty[W_M \qty(\ketbra{\psi} \otimes \ketbra{u_0})W_M^\dagger] =%
            \tr_\mathcal{H}\qty[\qty(W_M \ket{\psi \otimes u_0})\qty(W_M \ket{\psi \otimes u_0})^\dagger]\\
			&= \sum_{\ell = 1}^N \qty(\bra{\psi_\ell}\otimes I_\mathcal{G}) \qty(W_M \ket{\psi \otimes u_0})%
            \qty(W_M \ket{\psi \otimes u_0})^\dagger \qty(\ket{\psi_\ell}\otimes I_\mathcal{G})\\
			&= \sum_{\ell = 1}^N \sum_{k,j=1}^m\qty(\bra{\psi_\ell}\otimes \ketbra{u_k}) \qty(W_M \ket{\psi \otimes u_0}) %
            \qty(W_M \ket{\psi\otimes u_0})^\dagger \qty(\ket{\psi_\ell}\otimes \ketbra{u_j})\\
			&=\sum_{\ell = 1}^N \sum_{k,j=1}^m\qty(\matrixel{\psi_\ell\otimes u_k}{W_M}{\psi \otimes u_0} \ket{u_k})%
            \qty(\matrixel{\psi \otimes u_0}{W_M^\dagger}{\psi_\ell \otimes u_j} \bra{u_j})\\
			&=\sum_{\ell = 1}^N \sum_{k,j=1}^m\left( \matrixel{\psi_\ell \otimes u_k}{W_M}{\psi \otimes u_0}%
            \matrixel{\psi \otimes u_0}{W_M^\dagger}{\psi_\ell \otimes u_j}\right) \ketbra{u_k}{u_j}.
		\end{split}
		\]
		Now, by applying $\bra{u_k} \qty(\cdot) \ket{u_k}$ to both sides of the identity above and taking into account Eq.~\ref{eq_povm_stinespring_probabilidades}, it follows:
		\[
		\begin{split}
			\matrixel{u_k}{\tr_\mathcal{H}\qty[W_M \qty(\ketbra{\psi} \otimes \ketbra{u_0})W_M^\dagger]}{u_k} &%
			= \sum_{\ell=1}^N\matrixel{\psi_\ell;u_k}{W_M}{\psi;u_0}\hspace{-1.75mm}\matrixel{\psi;u_0}{W_M^\dagger}{\psi_\ell;u_k}\\%
            &=\sum_{\ell=1}^N\abs{\matrixel{\psi_\ell}{M_k}{\psi}}^2\\
			&=\mathrm{Pr}\qty(M_k\vert \psi).
		\end{split}
		\]
		\cqd
	\end{proof}
\end{proposition}

\begin{remark}
	Considering a POVM measurement as a \emph{random vector} in the state space $\mathcal{D}(\mathcal{H})$, we find that the action of the corresponding quantum channel (defined by Eq.~\ref{eq_povmchannel}) results precisely in the \emph{expected value state}:\footnote{%
		For a definition of a random vector and related statistical concepts, see \cite{Anderson2003}.
	}
	\begin{equation}\label{eq_povm_canal-quantico}
		\mathcal{E}_M [\rho] = \sum_{k=1}^{m} Pr(k\vert \rho) \rho_k. 
	\end{equation}
\end{remark}

\begin{remark}
	Proposition \ref{teo_povm_unitario-stinespring} provides expressions for calculating the probabilities of outcomes for a POVM measurement in terms of a unitary dilation operator. Essentially, this is the content of Naimark’s theorem (or Neumark’s theorem), which relates a POVM on a Hilbert space $\mathcal{H}$ to a projective measurement on a larger Hilbert space, such as $\mathcal{H}\otimes \mathcal{G}$ \cite{Bengtsson2017, Watrous2018}. Physically, this means that a POVM measurement on a quantum system can be interpreted as a projective measurement performed on the system coupled to an independently prepared ancilla quantum system, which is subsequently discarded \cite{Peres1990, Rabelo2006}.
\end{remark}

Below, we present an example of the simulation of measurement operators in a single-qubit system which ilustrates the application of Proposition \ref{teo_povm_unitario-stinespring}.

\subsection{Example of quantum simulation of a POVM measurement on a single qubit}

For this example, we consider the measurement on the single qubit system defined by the POVM $M$ having the following measurement operators:%
\footnote{%
	The positive operator-valued measure presented in this example was discussed in \cite{Pinto2023}.
}
\[
\begin{split}
	M_1 &= \frac{1}{2\sqrt{2}}\qty(\ketbra{0} + \sqrt{3}\ketbra{1}{0} + 2\ketbra{1}),\ %
	M_2 = \frac{1}{2\sqrt{2}}\qty(\ketbra{0} - \sqrt{3}\ketbra{1}{0} + 2\ketbra{1}).
\end{split}
\]
Since we have two measurement operators, the ancillary Hilbert space $\mathcal{G}$ has two dimensions and we fix the computational basis, as done in Section \ref{secao_prepest_exemplo}:
\[
\ket{u_1} = \ket{0}\ \text{and}\ \ket{u_2} = \ket{1}.
\]
The primary ancilla state and the corresponding unitary dilation $W_M: \mathcal{H}\otimes\mathcal{Q}\rightarrow \mathcal{H}\otimes\mathcal{Q}$ for the associated POVM quantum channel are, respectively:
\[
\ket{u_0} = \ket{0}
\]
and
\[
\begin{split}
	W_M\ket{j,k} = \ket{\phi_{jk}}\ (j,k = 0, 1),
\end{split}
\]
where
\[
\begin{split}
	& \ket{\phi_{00}} := V\ket{00} =\frac{\ket{00} + \sqrt{3}\ket{10} + \ket{01} - \sqrt{3}\ket{11}}{2\sqrt{2}},\\
	& \ket{\phi_{10}} := V\ket{10} = \frac{\ket{10} +\ket{11}}{\sqrt{2}},\\
	& \ket{\phi_{01}} := \frac{\ket{00} - \ket{01}}{\sqrt{2}},\\
	& \ket{\phi_{11}} := \frac{\sqrt{3}\ket{00} - \ket{10} + \sqrt{3} \ket{01} + \ket{11}}{2\sqrt{2}}.
\end{split}
\]
 Now, we focus on the system's statevector
\[
\ket{+} = \frac{\ket{0} + \ket{1}}{\sqrt{2}} = H \ket{0},\ \text{where}\ H = \frac{1}{\sqrt{2}}\begin{pmatrix}
	1 & 1\\
	1 & 1
\end{pmatrix}.
\]

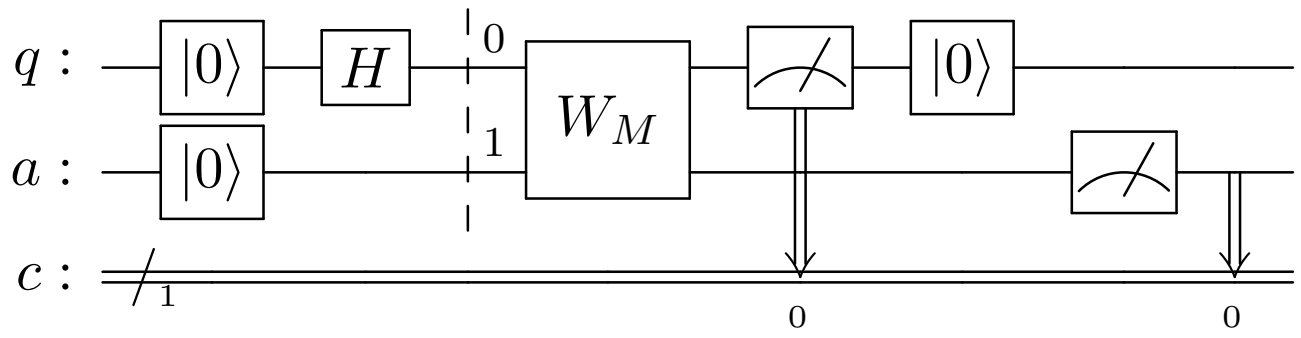
\begin{figure}
	\scalebox{1.0}{
		\Qcircuit @C=1.0em @R=0.2em @!R { \\
			\nghost{{q} :  } & \lstick{{q} :  } & \gate{\left|0\right\rangle} & \gate{H} \barrier[0em]{1} & \qw & \multigate{1}{W_M}_<<<{0} & \meter & \gate{\left|0\right\rangle} & \qw & \qw & \qw \\
			\nghost{{a} :  } & \lstick{{a} :  } & \gate{\left|0\right\rangle} & \qw & \qw & \ghost{W_M}_<<<{1} & \qw  & \qw & \meter & \qw & \qw\\
			\nghost{{c} :  } & \lstick{{c} :  } & \lstick{/_{_{1}}} \cw & \cw & \cw & \cw & \dstick{_{_{\hspace{0.0em}0}}} \cw \ar @{<=} [-2,0] & \cw & \cw & \dstick{_{_{\hspace{0.0em}0}}} \cw \ar @{<=} [-1,0] & \cw \\
			\\ }}
	\vspace{-5mm}
	\caption{\label{fig_povm+}Quantum circuit that simulates the measurement of the system in the state $\ket{+}$ by the POVM $M$. Here, $q$ (top) is the system qubit and $a$ (bottom) is the ancilla qubit.}
\end{figure}

To get the valures of $\mathrm{Pr}\qty(M_1 \vert +)$ and $\mathrm{Pr}\qty(M_2 \vert +)$ via quantum computational simulation, we can use Eq.~\ref{eq_povm_probabilidades2} of Proposition \ref{teo_povm_unitario-stinespring}, which due to our choice for the primary ancilla state is reduced to%
\footnote{
	Note that $X^\dagger = X$.
}
\[
\begin{split}
	\mathrm{Pr}\qty(M_1 \vert +) &= \abs{\matrixel{00}{W_M}{+0}}^2+ \abs{\matrixel{1 0}{W_M}{+ 0}}^2\\
	&\quad = \abs{\matrixel{0 0}{W_M\qty(H \otimes I)}{00}}^2  +   \abs{\matrixel{1  0}{W_M\qty(H \otimes I)}{0 0}}^2
\end{split}
\]
and
\[
\begin{split}
	\mathrm{Pr}\qty(M_2 \vert +) &= \abs{\matrixel{0 1}{W_M}{+0}}^2 +  \abs{\matrixel{1  1}{W_M}{+ 0}}^2\\
	&\quad = \abs{\matrixel{01}{W_M\qty(H \otimes I)}{0 0}}^2 +  \abs{\matrixel{1  1}{W_M\qty(H \otimes I)}{00}}^2
\end{split}
\]
Figure \ref{fig_povm+} shows the quantum circuit for simulating the aforementioned measurement, with the goal of obtaining $\mathrm{Pr}\qty(M_2 \vert +)$. The post-processing of this simulation consists of calculating the ratio between the number of results (counts) $1$ to the total number of runs (shots). To obtain $\mathrm{Pr}\qty(M_1 \vert +)$, the circuit and the post-processing are similar.

We simulated an experiment with this circuit using the \emph{AerSimulator} from the \emph{Qiskit} SDK with $2^{15}$ shots and obtained the following results:
\footnote{%
	GiHub page for the code:\newline \url{https://github.com/LucioFassarella/SQC/blob/main/SQC_povm.ipynb}.
}
\[
\begin{split}
	\mathrm{Pr}\qty(M_1 \vert +)_\text{sim} &\approx 0.933837890625,\\ %
	\mathrm{Pr}\qty(M_2 \vert +)_\text{sim} &\approx 0.066162109375.
\end{split}
\]

The theoretical probabilities are
\[
\begin{split}
	\mathrm{Pr}\qty(M_1 \vert +) &= \tr\qty(M_2 \ketbra{+} M_2^\dagger)%
	= \frac{1}{2} + \frac{\sqrt{3}}{4} \approx 0,933012702;
\end{split}
\]
\[
\begin{split}
	\mathrm{Pr}\qty(M_2 \vert +) &= \tr\qty(M_2 \ketbra{+} M_2^\dagger)%
	= \frac{1}{2} - \frac{\sqrt{3}}{4} \approx 0,066987298.
\end{split}
\]
As can be seen, the theoretical and simulated results are fully consistent.%

\section{\label{secao_sistemas-markovianos}Simulation of Markovian Quantum Systems}

Consider a quantum system with Hilbert space $\mathcal{H}$, and let $\{\ket{\psi_\ell}; \ell = 1, \dots, N\}$ be an orthonormal basis for $\mathcal{H}$:
\[
\mathcal{H} = \mathrm{span}\qty{\ket{\psi_\ell};\ \ell = 1,\ \dots, N}.
\]
We assume that the evolution of the system is Markovian, which means that its evolution is described by a \emph{quantum dynamical semigroup} $T_{\tau\ge 0}: \mathcal{L}\qty[\mathcal{H}]\rightarrow \mathcal{L}\qty[\mathcal{H}]$:
\[
\rho(t + \tau) =  T_{\tau}\qty[\rho(t)],\ \forall t,\tau \ge 0.
\]
The semigroup property reads 
\[
T_{\tau_2}T_{\tau_1} = T_{\tau_2 + \tau_1},\ \forall \tau_1, \tau_2 \ge 0,
\]
and implies that the system's state at any given instant depends only on the immediately preceding state, and not on its past history. In such a case, the system's dynamics is characterized by the \emph{Lindblad equation} \cite{Alicki1987, Breuer2002, Gorini1976, Lindblad1976, Lidar2020}:
\begin{equation}\label{eq_lindblad}
	\begin{split}
		&\frac{d}{dt}\rho(t) = \mathfrak{L}_t[\rho(t)],
	\end{split}
\end{equation}
where $\mathfrak{L}_t: \mathcal{L}(\mathcal{H}) \to \mathcal{L}(\mathcal{H})$ is the termed \emph{Liouvillian superoperator} and has the general form
\[
\mathfrak{L}_t[\cdot] = -\frac{i}{\hbar} [H, \cdot] + \sum_{k = 1}^m\gamma_k\left(L_k \cdot L_k^\dagger - \frac{1}{2}\{L_k^\dagger L_k, \cdot\}\right);
\]
here $\gamma_k \ge 0$ are the \emph{transition} (or \emph{decay}) \emph{rates} (with dimensions of inverse time), and $L_k: \mathcal{H} \to \mathcal{H}$ are the (dimensionless) \emph{Lindblad operators}, while $[\cdot, \cdot]$ and $\{\cdot, \cdot\}$ denote the commutator and anticommutator, respectively. The Hamiltonian, Lindblad operators, and transition rates can be time-dependent; however, we will not explicitly denote this dependence to simplify the notation.%
\footnote{%
	When the Liouvillian is time-independent, the \emph{quantum dynamical semigroup} is given by the exponential of the Liouvillian: $T_\tau = e^{\tau\mathfrak{L}},\ \forall \tau\ge 0$.
}

The Lindblad equation allows us to write the following linear approximation for the evolution of a quantum state over an infinitesimal interval $dt$:
\begin{equation}\label{eq_lindbladinf}
	\rho(t + dt) = \rho(t) + \mathfrak{L}\qty[\rho(t)]dt + O(dt^2)
\end{equation}
This infinitesimal evolution corresponds to the action of a quantum channel $\mathcal{E}_{dt}$ on $\mathcal{H}$, since it must map states to states, be linear trace-preserving and completely positive -- from what we might term an \emph{infinitesimal evolution channel}.This channel is a first-order approximation of the quantum dynamical semigroup, which allows us to construct a quantum simulation of the evolution by applying developments presented in Section \ref{secao_canaisq}. For this purpose, we need to construct a Choi-Kraus representation for $\mathcal{E}_{dt}$; in other words, we must define \emph{infinitesimal Kraus operators} such that:
\footnote{To simplify subsequent notation, we note that the index for the Kraus operators in the summation begins at $j=0$, whereas the summation in the Lindblad superoperator begins at $k=1$.}
\begin{equation}\label{eq_kraus}
	\begin{split}
		\mathcal{E}_{dt}\qty[\rho(t)] = \rho(t+dt) %
		=\sum_{j=0}^s K_j(dt)\rho(t)K_j^\dagger(dt).
	\end{split}
\end{equation}
Identifying the expressions of Eq.~\ref{eq_lindbladinf} and Eq.~\ref{eq_kraus}, we obtain
\begin{equation}\label{eq_lindbladinf-kraus}
	\begin{split}
		\sum_{j=0}^s K_j\qty(dt)\rho(t)K_j\qty(dt)^\dagger &= \rho(t) -\frac{i}{\hbar} \qty[H,\rho(t)]dt +\\
		&+\sum_{k=1}^m\gamma_k\qty(L_k\rho(t)L_k^\dagger - \frac{1}{2}\qty{L_k^\dagger L_k ,\rho(t)})dt%
		+ O(dt^2).
	\end{split}
\end{equation}
Following \cite[Chapter IX, pp.45-46]{Lidar2020} (or the related \cite[Eq.6]{Ding2024}), we take $s = m$ and define  directly from Lindblad operators the new operator
\[
L_0 := -\frac{1}{2}\sum_{j=1}^m \gamma_jL_j^\dagger L_j- \frac{i}{\hbar} H
\] 
and the \emph{infinitesimal Kraus operators}
\begin{equation}\label{eq_kraus_def} 
	\begin{split}
		K_0\qty(dt) &:= I + L_0 dt + R dt^2;\ %
		K_j\qty(dt) := \sqrt{\gamma_j dt}L_j ,\ j = 1, \dots, m,
	\end{split}
\end{equation}
where we have introduced the \emph{residual operator} $R = R(dt)$ to ensure afterwards the validity of the \emph{isometry condition}:
\begin{equation}\label{eq_isometry1}
	\sum_{j=0}^m K_j\qty(dt)^\dagger K_j\qty(dt) = I.
\end{equation}
By direct inspection, we verify that the operators defined by Eqs. \ref{eq_kraus_def} indeed establish a Choi-Kraus representation for the infinitesimal evolution to first order:
\[
\begin{split}
	&\sum_{j=0}^m K_j\rho(t)K_j^\dagger =
	\qty(I+ L_0dt + R dt^2)\rho(t)\qty(I+ L_0^\dagger dt + R^\dagger dt^2) +%
    \sum_{j=1}^m \sqrt{\gamma_j dt} L_j\rho(t)\sqrt{\gamma_j dt}L_j^\dagger\\
	&= \rho(t) + (L_0\rho(t) + \rho(t) L_0^\dagger )dt + \sum_{j=1}^m \gamma_jL_j \rho(t) L_j^\dagger  dt+\\%
	&\quad +\qty(L_0\rho(t) L_0^\dagger + R\rho(t) + \rho(t) R^\dagger)dt^2+%
    \qty(L_0\rho(t) R^\dagger + R \rho(t) L_0^\dagger)dt^3 + R\rho(t) R^\dagger dt^4\\
	&= \rho(t) -\frac{i}{\hbar} \qty[H,\rho(t)]dt+%
    \sum_{j=1}^m \gamma_j\qty(L_j \rho(t) L_j^\dagger  - \frac{1}{2}\qty{L_j^\dagger L_j,\rho(t)})dt+\\%
	&\quad + \qty(L_0\rho(t) L_0^\dagger + R\rho(t) + \rho(t) R^\dagger)dt^2+%
    \qty(L_0\rho(t) R^\dagger + R \rho(t) L_0^\dagger)dt^3+%
    R\rho(t) R^\dagger dt^4\\
	& = \rho(t) + \mathfrak{L}\qty[\rho(t)]dt +\\
	&\quad + \qty(L_0\rho(t) L_0^\dagger + R\rho(t) + \rho(t) R^\dagger)dt^2 +%
    \qty(L_0\rho(t) R^\dagger + R \rho(t) L_0^\dagger)dt^3+%
    R\rho(t) R^\dagger dt^4;
\end{split}
\]
that is
\begin{equation}\label{eq_evolucao_kraus}
	\begin{split}
		&\sum_{j=0}^m K_j\rho(t) K_j^\dagger  = \rho(t) + \mathfrak{L}\qty[\rho(t)]dt + \mathcal{R}_{dt}\qty[\rho(t)]dt^2,
	\end{split}
\end{equation}
where we have introduced the \emph{residual superoperator}:
\[
\begin{split}
	\mathcal{R}_{dt}\qty[\Gamma] &= \qty(L_0\Gamma L_0^\dagger + R\Gamma + \Gamma R^\dagger) +%
    \qty(L_0\Gamma R^\dagger + R \Gamma L_0^\dagger)dt+%
    RR^\dagger dt^2,\ \forall \Gamma \in \mathcal{L}\qty(\mathcal{H}).
\end{split}
\]
Similarly, we calculate the expression for the isometry condition, Eq.~\ref{eq_isometry1}:%
\footnote{This equation can be derived in a manner analogous to the calculation of the identity in Eq.~\ref{eq_evolucao_kraus}. It can also be obtained from that identity by setting $\rho = I$ and swapping the positions of the operators and their respective adjoints (that is, by strictly reversing the order of the non-Hermitian operators and their respective adjoints in each monomial of the expansion).
}
\begin{equation}\label{eq_isometry2}
	\begin{split}
		\sum_{j=0}^m K_j^\dagger K_j &= I + \qty[ \qty(L_0^\dagger L_0 + R + R^\dagger) \right. +
		\left. \qty(L_0^\dagger R + R^\dagger L_0)dt + R^\dagger R dt^2] dt^2.
	\end{split}
\end{equation}
Thus, the \emph{isometry condition} on the infinitesimal Kraus operators is equivalent to the fact that the residual operator $R$ satisfies the following \emph{residual equation}:
\begin{equation}\label{eq_residual}
	\begin{split}
		\qty(L_0^\dagger L_0 + R + R^\dagger) + \qty(L_0^\dagger R + R^\dagger L_0)dt + R^\dagger R dt^2 = 0.
	\end{split}
\end{equation}
In summary, the Kraus operators defined by Eq.~\ref{eq_kraus_def}, with the residual operator $R$ satisfying the residual equation Eq.~\ref{eq_residual}, actually define a quantum channel that corresponds to the infinitesimal evolution $\mathcal{E}_{dt}$ to first order in time 

Now, let $W_{dt}: \mathcal{H}\otimes \mathcal{G}\rightarrow \mathcal{H}\otimes \mathcal{G}$ denote the unitary dilation corresponding to the quantum channel $\mathcal{E}_{dt}$ with a primary ancilla state $\ket{u_0} \in \mathcal{G}$; that is,
\[
\begin{split}
	\rho(t + dt) &= \tr_{\mathcal{G}}\qty(W_{dt}\qty(\rho(t)\otimes\ketbra{u_0})W_dt^\dagger) + O\qty(dt^2).
\end{split}
\]
By induction, we obtain the recursive expression for all $M \in \mathbb{N}$:
\begin{equation}\label{eq_evolucao}
	\begin{split}
		\rho(t + (M+1)dt) &=\tr_{\mathcal{G}}\qty(W_{dt} \qty(\rho(t+ Mdt) \otimes \ketbra{u_0})W_{dt}^\dagger) + O\qty(dt^2).
	\end{split}
\end{equation}

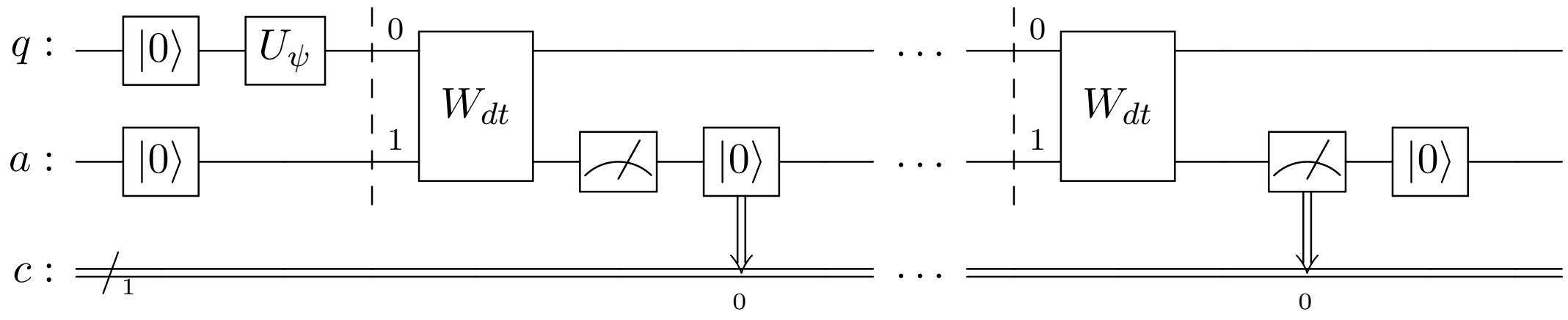
\begin{figure}
	\scalebox{1.0}{
		\Qcircuit @C=1.1em @R=1.0em @!R { \\
			\nghost{{q} : } & \lstick{{q} : } &\gate{\left|0\right\rangle} &\gate{U_\psi} 
			\barrier[0em]{1} &\qw 
			&\multigate{1}{W_{dt}}_<<<{0} &\qw &\qw &\qw &\qw
			&\dots &
			\barrier[0em]{1} &\qw
			&\multigate{1}{W_{dt}}_<<<{0} &\qw &\qw &\qw
			& \qw & \qw\\
			\nghost{{a} :  } & \lstick{{a} :  } &\gate{\left|0\right\rangle} &\qw &\qw
			&\ghost{W_{dt}}_<<<{1} &\meter &\gate{\left|0\right\rangle} &\qw &\qw
			&\dots &
			&\qw  &\ghost{W_{dt}}_<<<{1}  &\qw & \meter &\gate{\left|0\right\rangle} &\qw &\qw\\
			\nghost{{c} :  } & \lstick{{c} :  } & \lstick{/_{_{1}}} \cw &\cw &\cw &\cw &\cw
			&\dstick{_{_{\hspace{0.0em}0}}} \cw \ar @{<=} [-1,0] &\cw &\cw & \dots & &\cw &\cw &\cw &\dstick{_{_{\hspace{0.0em}0}}} \cw \ar @{<=} [-1,0] &\cw &\cw &\cw\\
			\\ }}
	\caption{\label{fig_timev}Quantum circuit for simulating the evolution of a Markovian quantum system from an initial state $\ket{psi} = U_\psi \ket{0}$, according with Eq.~\ref{eq_evolucao}. Here, $q$ is the set of qubits describing the system, $a$ the set of ancillary qubits, $U_\psi$ the gate that prepares initial state and $W_{dt}$ is the unitary dilation for the evolution quantum channel at a first-order. The number of unitary dilations corresponds to the number $M$ of simulation steps. $U_\psi$ can be replaced by the procedure from Section \ref{secao_prepest} for preparing an initial mixed state $\rho$.}
\end{figure}

Figure \ref{fig_timev} presents the general form of the quantum circuit for the finite evolution of a Markovian quantum system we cand built from Eq.~\ref{eq_evolucao}.

\subsection{Computational complexity}

\textbf{Step-scaling.} The first-order approximation Eq.~\ref{eq_lindbladinf-kraus} has a local error of $O(dt^2)$ per step. Accumulating this over $M = T/dt$ steps, the global error is
\[
M \cdot O(dt^2) = O(T \cdot dt) = O(T^2/M).
\]
Therefore, to achieve a precision $\varepsilon$ in a simulation up to time $T$, one requires
\[
M = O(T^2/\varepsilon).
\]
This step-scaling corresponds precisely to the behavior of a first-order Trotter-Suzuki formula \cite{Nielsen2010}, which we have extend here from unitary evolution to a general Markovian evolution.

Taking into account the gate cost for the UD strategy for the simulation of a composition of $M$ quantum channels given by Eq.~\ref{cc_totalqubits}, the total number of two-qubit has order
\[
O\qty(N^2m^2T^2/\varepsilon).
\]

In Subsection \ref{secao_sqm_exemplo}, we present an example of the quantum simulation of Eq.~\ref{eq_evolucao} on a digital quantum computer. Next, we offer some considerations regarding the simulation of closed systems, whose dynamics are governed by the Schrödinger equation. 

\subsection{Simulation of the evolution of closed quantum systems}\label{secao_sqf}

In this section, we analyze the application of our simulation strategy to the special case of closed quantum systems 

The evolution of a closed quantum system is generally determined exclusively by the system's Hamiltonian, say $\mathfrak{h}$, according with Schrödinger equation:
\[
i\hbar \frac{d }{dt}\ket{\psi(t)} = \mathfrak{h}\ket{\psi(t)}.
\]
This equation is a special case of the Lindblad equation Eq.~\ref{eq_lindblad}, in which the Liouvillian reduces to the Hamiltonian term and there are no Lindblad operators. Therefore, there is only one infinitesimal Kraus operator:
\[
K_0(dt) = I - \frac{i}{\hbar}\mathfrak{h}dt + Rdt^2.
\]
The residual equation for the residual operator $R=R\qty(dt)$ is reduced to
\[
\frac{1}{\hbar^2}\mathfrak{h}^2 + R + R^\dagger + \frac{i}{\hbar}\qty(\mathfrak{h}R - R^\dagger \mathfrak{h})dt + R^\dagger Rdt^2 = 0.
\]
The isometry condition Eq.~\ref{eq_isometry1} implies that $K_0$ is unitary, so the corresponding infinitesimal evolution channel is unitary too:
\[
\mathcal{E}_{dt}(\Gamma) := K_0 \Gamma K_0^\dagger.
\] 
In fact, $K_0$ is a first order approximation for the evolution operator $e^{-i\mathfrak{h}t/\hbar}$:
\[
\begin{split}
	\ket{\psi(t+dt)} &= \ket{\psi(t)} -\frac{i}{\hbar}\mathfrak{h}\ket{\psi(t)}dt + O(dt^2)%
	=K_0 \ket{\psi(t)} +O(dt^2)
\end{split}
\]
Therefore, to simulate the evolution of the system, it suffices to apply the following recurrence:
\begin{equation}\label{eq_simula-evolucao-unitaria}
	\ket{\psi(t+(M+1)dt)} = K_0\ket{\psi(t +Mdt)},\ \forall M = 0, 1, \dots.
\end{equation}
\begin{remark}
	The recurrence relation provided by Eq.~\ref{eq_simula-evolucao-unitaria} does not require the construction of a unitary dilation, nor does it depend on the Trotterization of the system's evolution operator -- a step typically required by the standard approach \cite{Nielsen2010}.
\end{remark}
\subsection{Example of quantum simulation of the evolution of a quantum system}\label{secao_sqm_exemplo}

We consider the single qubit with amplitude damping, for which the Lindblad equation is given by%
\footnote{
	\cite{Wang2026} studied a similar model, using a different approach to simulate its evolution.
}
\[
\frac{d\rho}{dt} = -\frac{i}{\hbar}\qty[\mathfrak{h},\rho] + \gamma \qty[\sigma^+ \rho(t)\sigma^- - \frac{1}{2}\qty{\sigma^-\sigma^+,\rho(t)}],
\]
where we have the Hamiltonian and ladder operators
\[
\mathfrak{h} =  -\epsilon Z,\ \sigma^+ = \ketbra{0}{1},\ \ \sigma^- = \ketbra{1}{0}.
\]
In this case, the system's Liouvillian has only one Lindblad operator, namely,
\[
\begin{split}
	L_1 &:= \sigma^+.
\end{split}
\]
The exact solution for the equation of motion, derived in Appendix \ref{appendix_b}, is given by
\begin{equation}\label{eq_tev_state}
	\begin{split}
		\rho(t) &= \begin{pmatrix}
			1 - e^{-\gamma t}a & e^{(-2i\epsilon/\hbar - \gamma/2)t}b\\
			\ \\
			e^{(2i\epsilon/\hbar - \gamma/2)t}\bar{b} & e^{-\gamma t}a
		\end{pmatrix},
	\end{split}
\end{equation}

where
\[
a \in \mathbb{R},\ b \in \mathbb{C},\ (1- a)a \ge |b|^2.
\]
From this, we derive the exact evolution of the energy expectation value:
\begin{equation}\label{eq_tev_energy}
	\begin{split}
		\langle \mathfrak{h} \rangle_\rho (t) = tr\qty(\rho\qty(t) \mathfrak{h})
		= (2e^{-\gamma t}a - 1)\epsilon.
	\end{split}
\end{equation}
For our computational simulation, we fix a time step $dt$ and follow previous developments to define the infintesimal Kraus operators from Eq.~\ref{eq_kraus_def}:
\[
\begin{split}
	K_0 &:= I +\qty(-\frac{\gamma}{2}\vert 1\rangle\langle 1 \vert + i\tilde{\epsilon} Z )dt + Rdt^2,\\
	K_1 &:= \sqrt{\gamma dt}\sigma^+ = \sqrt{\gamma dt}\ketbra{0}{1}.
\end{split}
\]
where
\[
L_0 := - \frac{\gamma}{2} L_1^\dagger L_1 - \frac{i}{\hbar}H = - \frac{\gamma}{2}\vert 1\rangle\langle 1 \vert + i\tilde{\epsilon} Z
\]
and
\[
\tilde{\epsilon}:= \frac{\epsilon}{\hbar}.
\]

\begin{figure}
	\centering
	\includegraphics[width=1.0\linewidth]{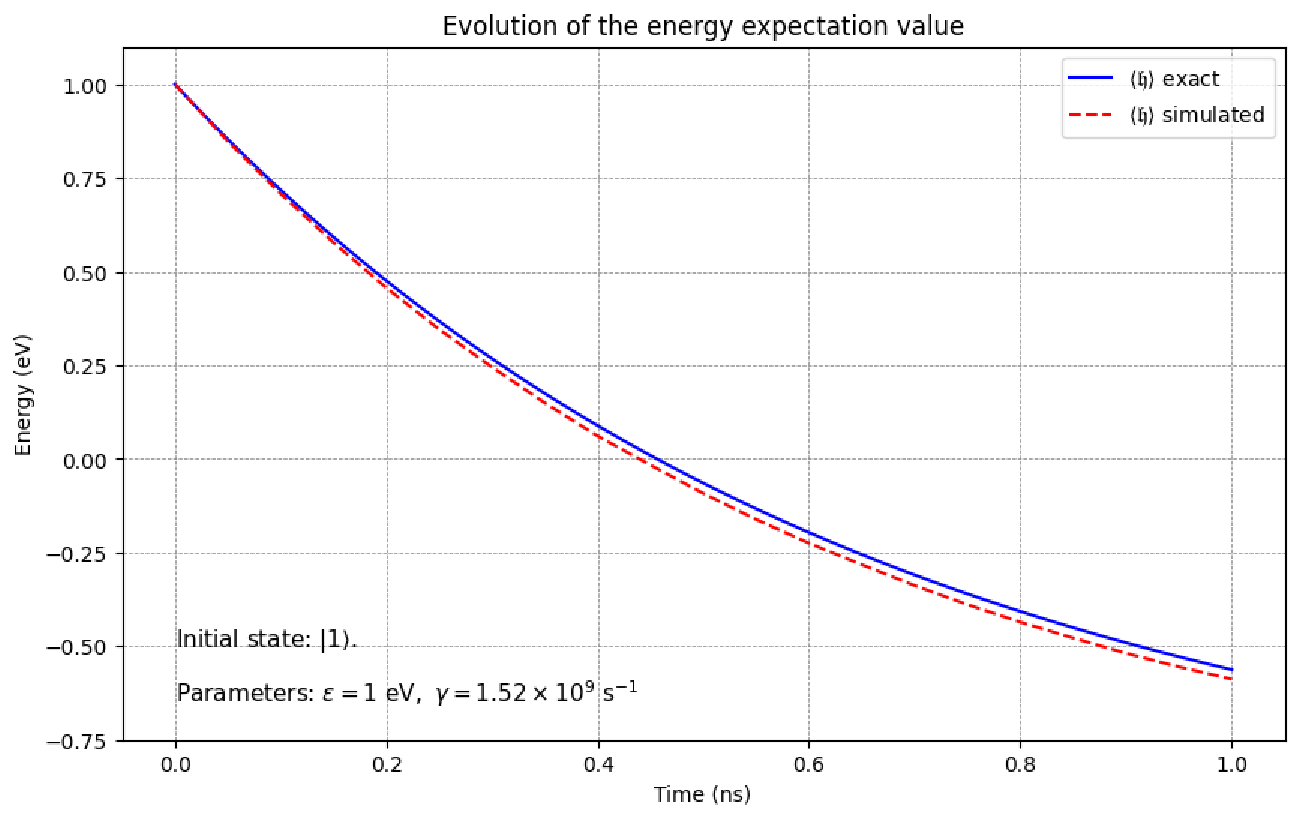}
	\caption{\label{fig_energia}Exact and simulated evolution of the energy expectation value of the qubit with amplitude damping.}
	
\end{figure}

The residual equation Eq.~\ref{eq_residual} for the residual operator $R = R\qty(dt)$ takes the form
\[
\begin{split}
	&\frac{\gamma^2}{4}\vert 1\rangle\langle 1 \vert - \tilde{\epsilon}^2I + R + R^\dagger - \frac{\gamma}{2}( \vert 1\rangle\langle 1 \vert R + R^\dagger \vert 1\rangle\langle 1 \vert + %
	i\tilde{\epsilon} ZR^\dagger - i\tilde{\epsilon} R Z )dt + R^\dagger R dt^2 = 0.
\end{split}
\]
With the anzats
\[
R = r_0 \vert 0\rangle\langle 0 \vert + r_1 \vert 1\rangle\langle 1 \vert\ \ (r_0,r_1 \in \mathbb{R})
\]
the residual equation is reduced to a pair of quadratic equations for the coefficients $r_0$ and $r_1$:
\[
\begin{split}
	&0 = \frac{\gamma^2}{4}\vert 1\rangle\langle 1 \vert + \tilde{\epsilon}^2 \vert 0\rangle\langle 0 \vert + \tilde{\epsilon}^2\vert 1\rangle\langle 1 \vert + 2r_0 \vert 0\rangle\langle 0 \vert + \\
	& \quad \quad +  2r_1\vert 1\rangle\langle 1 \vert - \gamma r_1 dt \vert 1\rangle\langle 1 \vert +\\
	&\quad \quad + dt^2r_0^2\vert 0\rangle\langle 0 \vert + dt^2r_1^2\vert 1\rangle\langle 1 \vert\\
	&\ = \left(\tilde{\epsilon}^2  + 2r_0 + dt^2r_0^2 \right) \ketbra{0} +\\
	&\quad \quad +
	\left(\frac{\gamma^2}{4} + \tilde{\epsilon}^2 + 2b - \gamma dt b + dt^2b^2\right)\ketbra{1}\\
	&\iff \left\lbrace
	\begin{array}{l}
		\tilde{\epsilon}^2  + 2r_0 + dt^2r_0^2 = 0, \\
		\frac{\gamma^2}{4} + \tilde{\epsilon}^2 + (2 - \gamma dt) r_1 + dt^2r_1^2 = 0.
	\end{array}
	\right.
\end{split}
\]
Explicit solutions for $r_0$ and $r_1$:
\[
\begin{split}
	r_0& = - \frac{1 - \sqrt{1- \tilde{\epsilon}^2dt^2}}{dt^2} %
	= - \frac{1}{2}\tilde{\epsilon}^2 + O(dt^2)
\end{split}
\]
and
\[
\begin{split}
	r_1  &= - \frac{1 - \gamma dt/2 - \sqrt{1- \gamma dt - \tilde{\epsilon}^2 dt^2}}{dt^2} %
	 = - \left( \frac{\tilde{\epsilon}^2}{2} + \frac{\gamma^2}{8} \right) + O(dt^2).
\end{split}
\]
The image of the Stinespring isometry is
\[
\begin{split}
	\mathcal{V} &= \mathrm{span}\left\{K_0\ket{00} + K_1\ket{01},\ %
	K_0\ket{10} + K_1\ket{11} \right\}\\
	&=\mathrm{span}\left\{\kappa_{0}\ket{00},\ %
	\kappa_{1} \ket{10} + \kappa_{2}\ket{01} \right\},
\end{split}
\]
where
\[
\begin{split}
	\kappa_{0} &:= 1 + r_0dt^2 + i\epsilon dt\\
	\kappa_{1} &:= 1 - \gamma dt/2  + r_1dt^2 - i\epsilon dt\\
	\kappa_{2} &:= \sqrt{\gamma dt}.
\end{split}
\]
To define the unitary dilation, we fix the following orthonormal basis for $\mathcal{V}$ and for $\mathcal{V}^\perp$, respectively:
\[
\begin{split}
	\vert\phi_{00}\rangle &:= \kappa_{0}\vert 0\rangle\otimes\vert 0\rangle \\
	\vert\phi_{01}\rangle &:= +\kappa_{1} \vert 1\rangle\otimes\vert 0\rangle + \kappa_{2}\vert 0\rangle\otimes\vert 1\rangle,
\end{split}
\]
and
\[
\begin{split}
	\vert\phi_{10}\rangle &:=- \kappa_{2} \vert 1\rangle\otimes\vert 0\rangle + \kappa_{1}\vert 0\rangle\otimes\vert 1\rangle,\\
	\vert\phi_{11}\rangle &:= \kappa_{0}\vert 1\rangle\otimes\vert 1\rangle.
\end{split}
\]
Finally, we define the unitary dilation by its action on the computational basis, according with Eq.~\ref{eq_unitario_stinespring}:
\[
\begin{split}
	W_{dt}\qty(\vert j\rangle\otimes\vert k\rangle) := \vert \phi_{jk}\rangle,\ \ j,k=0,1.
\end{split}
\]
We simulated the evolution of this system using the \emph{Qiskit} SDK's \emph{AerSimulator}, performing $10^6$ executions (\textit{shots}) for each step of the simulation.%
\footnote{
	GiHub page for the code:\newline \url{https://github.com/LucioFassarella/SQC/blob/main/SQC_qubit_amplitude-dampping.ipynb}.
} %
 Figure \ref{fig_energia} shows the corresponding evolution of the expectation value of the energy, with system parameters $\epsilon = 1\ \mathrm{eV}$ and $\gamma = 1.52 \times 10^9\ s^{-1}$, initial state $\ket{\psi_0} = \ket{1}$ (corresponding to $a = 1$ and $b = 0$ in Eq.~\ref{eq_tev_state}), simulation duration of $1$ nanosecond, and a total of $20$ steps. As can be observed, the simulated results show reasonable agreement with the theoretical result. The behevior of the erros is also reasonable, as it grows with the simulation steps.

\section{\label{secao_conclusao}Conclusion}

Digital quantum computers operate according to the circuit model, processing states exclusively through unitary operators (quantum gates) and measurements. However, many physically relevant processes are non-unitary, including the preparation of mixed states, generalized measurements (POVMs), and the dynamics of open quantum systems. Here, we present a self-contained strategy -- which we term UD strategy -- to simulate such processes by using the Choi-Kraus operator-sum representation of a quantum channel to construct a unitary operator (called \emph{unitary dilation}) hat implements the channel within a circuit involving an auxiliary register (ancilla). To illustrate this, we apply the same construction to the aforementioned processes.

For the preparation of mixed states, we construct a channel based on the spectral decomposition of the density operator and apply our simulation strategy to this channel to prepare the mixed state. It is worth noting that the resulting unitary dilation operator acts as a purification of this state. An example involving a single qubit is presented, in which the expected value of an observable with respect to a mixed state is estimated via quantum simulation, with the results being in agreement with theoretical calculations.

In the case of POVMs, we define a corresponding quantum channel in the Choi-Kraus representation by identifying the Kraus operators with the POVM measurement operators. Our construction recovers the Naimark dilation and provides explicit expressions for outcome probabilities in terms of the unitary dilation. Here too, a POVM measurement of a single qubit is simulated, yielding results that are in agreement with theoretical calculations.

For the simulation of Markovian quantum dynamics, we introduce an operator, referred to as a \emph{residual operator}, that allows for the construction of the \emph{infinitesimal evolution channel} directly from the system's Liouvillian, which approximates the system's quantum dynamical semigroup to first order. This enables us to develop a quantum simulation of the dynamics in a manner analogous to Euler's method for ordinary differential equations. The method naturally encompasses unitary evolutions as special cases and completely circumvents Trotterization of the evolution operator for the purpose of quantum computational simulation. We illustrate the method by simulating the evolution of a single qubit subjected to amplitude damping, verifying the result by comparing it with the exact solution.

In our view, the main contribution of this work lies precisely in the concept of a \emph{residual operator}, which allows us to quantum-simulate the evolution of Markovian quantum systems within the UD strategy. The new burden lies in the need to solve the \emph{residual equation}, a quadratic equation for the residual operator which is required to obtain the trace-preserving property for the infinitesimal evolution channel. In the example we present, this equation was solved by introducing an \textit{ansatz} for the residual operator, reducing the problem to a pair of real quadratic equations. We believe it is worth studying this equation, with the goal of developing a general solution method suitable for application to the quantum simulation of Markovian quantum systems, thereby reducing pre-quantum processing calculations.

We also present an analysis of resources for the UD strategy, in terms of the number of qubits and the decomposition of gates into one- or two-qubit gates, as well as the step-scaling for the quantum simulation of the evolution of Markovian quantum systems. This analysis considers the worst-case scenario, leaving the exploration of efficiency improvements for future studies -- perhaps through combinations with other techniques, such as the \emph{linear combination of unitaries} (defined in \cite{Kothari2014} and applied in \cite{Cleve2019}).

Finally, all theoretical results are proven, with the exception of the Choi-Kraus theorem, for which we provide standard references. The simulations were performed using IBM's Qiskit AerSimulator, and the code is publicly available on GitHub.\\

\begin{acknowledgments}
I thank Gabriel Lourenço da Silva for discussions that led me to write this text.
\end{acknowledgments}

\appendix

\section{evolution of amplitude damping qubit}\label{appendix_b}

Lindblad equation:
\[
\begin{split}
	\frac{d\rho}{dt} &= -\frac{i}{\hbar} \left[H, \rho\right] + \gamma \left\lbrack\sigma^+\rho\sigma^- - \frac{1}{2}\left\{\sigma^-\sigma^+,\rho\right\}\right\rbrack;\\
	H &= -\epsilon Z = \begin{pmatrix}
		-\epsilon & 0 \\
		0 & \epsilon
	\end{pmatrix};\\
	\sigma^+ &= \begin{pmatrix}
		0 & 1 \\
		0 & 0
	\end{pmatrix}; \ \sigma^- = \begin{pmatrix}
		0 & 0 \\
		1 & 0
	\end{pmatrix}.
\end{split}
\]
Writing
\[
\tilde{\epsilon}:= \frac{\epsilon}{\hbar}
\]
and the density matrix in components
\[
\rho = \begin{pmatrix}
	a & b\\
	c & d
\end{pmatrix},
\]
we obtain the following relations
\[
\begin{split}
	i \left[H, \rho \right] &= \begin{pmatrix}
		0 & -2i\tilde{\epsilon} b\\
		2i\tilde{\epsilon} c & 0
	\end{pmatrix};\ 
	\gamma \sigma^+ \rho \sigma^- = \begin{pmatrix}
		\gamma d & 0\\
		0 & 0
	\end{pmatrix};\ 
	- \frac{\gamma}{2}\left\{\sigma^-\sigma^+,\rho\right\} = \begin{pmatrix}
		0 & -\gamma b/2\\
		-\gamma c/2 & -\gamma d
	\end{pmatrix}.
\end{split}
\]
The equation of motion becomes
\[
\frac{d\rho}{dt} = \begin{pmatrix}
	\dot{a} & \dot{b}\\
	\dot{c} & \dot{d}
\end{pmatrix} =%
\begin{pmatrix}
	\gamma d & (-2\tilde{\epsilon} i - \gamma /2)b\\
	(2\tilde{\epsilon} i - \gamma/2)c & -\gamma d
\end{pmatrix}.
\]
Therefore
\[
\left\lbrace
\begin{array}{ll}
	\dot{a} &= \gamma d,\\
	\dot{b} &= (- 2\tilde{\epsilon} i - \gamma/2)b,\\
	\dot{c} &= (2\tilde{\epsilon} i - \gamma/2)c,\\
	\dot{d} &= -\gamma d;
\end{array}
\right.
\]
and so, for some $\alpha \in \mathbb{C}$:
\[
\left\lbrace
\begin{array}{ll}
	a &= \alpha - e^{-\gamma t}d_0,\\
	b &= e^{(- 2\tilde{\epsilon} i - \gamma/2)t}b_0,\\
	c &= e^{(2\tilde{\epsilon} i - \gamma/2)t}c_0,\\
	d &= e^{-\gamma t}d_0.
\end{array}
\right.
\]
The following conditions must hold for any state:
\[
\begin{split}
	tr(\rho) = 1 &\iff \alpha = 1;\\
	\rho^\dagger = \rho &\iff \overline{e^{(-2\tilde{\epsilon} i - \gamma/2)t}b_0} = e^{(2\tilde{\epsilon} i - \gamma/2)t}c_0
	\iff \bar{b}_0 = c_0;\\
	\rho\ge 0 &\iff \det(\rho) = ad - bc \ge 0
	\Longleftarrow a_0d_0 - b_0c_0 \ge 0.
\end{split}
\]
So, the general solution is
\[
\rho(t) = \begin{pmatrix}
	1 - e^{-\gamma t}d_0 & e^{(-2i\tilde{\epsilon} - \gamma/2)t}b_0\\
	e^{(2i\tilde{\epsilon} - \gamma/2)t}\bar{b}_0 & e^{-\gamma t}d_0
\end{pmatrix},
\]
where
\[
d_0 \in \mathbb{R},\ b_0 \in \mathbb{C},\ (1- d_0)d_0 \ge | b_0|^2.
\]

\bibliographystyle{apsrev4-2}
\bibliography{bibliography}

\end{document}